\documentclass[draft]{agujournal2019}
\usepackage{url} 
\usepackage{lineno}
\usepackage[finalnew]{trackchanges} 
\usepackage{soul}
\newcommand{\corr}[1]{\textcolor{black}{#1}} 
\draftfalse

\journalname{JGR: Earth Surface}

\begin{document}

%
%

\title{Barchan-barchan \corr{dune} repulsion investigated at the grain scale}

\textcolor{blue}{An edited version of this paper was published by AGU. Copyright (2024) American Geophysical Union. Lima, N.C., Assis, W.R., Alvarez, C.A., Franklin, E.M., Barchan-barchan dune repulsion investigated at the grain scale. Journal of Geophysical Research: Earth Surface, 129, e2024JF007741, 2024\\
To view the published open abstract, go to https://doi.org/10.1029/2024JF007741.}

%
%




\authors{N. C. Lima\affil{1}, W. R. Assis\affil{1,3}, C. A. Alvarez\affil{2}, E. M. Franklin\affil{1}}


\affiliation{1}{Faculdade de Engenharia Mec\^anica, Universidade Estadual de Campinas (UNICAMP),\\
Rua Mendeleyev, 200, Campinas, SP, Brazil}

\affiliation{2}{Department of Earth \& Planetary Sciences, Stanford University,\\
Stanford, CA 94305, USA}

\affiliation{3}{Current address: Saint Anthony Falls Laboratory, University of Minnesota, Minneapolis, Minnesota, USA}




\correspondingauthor{Erick M. Franklin}{erick.franklin@unicamp.br}




\begin{keypoints}
\item We investigate numerically both the fluid flow and motion of grains for understanding the mechanisms behind barchan-barchan \corr{dune} repulsion
\item We measure the flow rate of grains and show that there is, indeed, greater erosion on the downstream barchan
\item The disturbed flow impacting the downstream barchan induces higher erosion (of its grains) and low accumulation (of upstream grains) 
\end{keypoints}

%
%

%
%


\begin{abstract}
Barchans are \corr{eolian} dunes of crescent shape found on Earth, Mars and other celestial bodies. Among the different types of barchan-barchan interaction, there is one, known as chasing, in which the dunes remain close but without touching each other. In this paper, we investigate the origins of this barchan-barchan \corr{dune} repulsion by carrying out grain-scale numerical computations in which a pair of granular heaps is deformed by the fluid flow into barchan dunes that interact with each other. In our simulations, data such as position, velocity and resultant force are computed for each individual particle at each time step, allowing us to measure details of both the fluid and grains that explain the repulsion. We show the trajectories of grains, time-average resultant forces, and mass balances for each dune, and that the downstream barchan shrinks faster than the upstream one, keeping, thus, a relatively high velocity although in the wake of the upstream barchan. In its turn, this fast shrinkage is caused by the flow disturbance, which induces higher erosion on the downstream barchan and its circumvention by grains leaving the upstream dune. Our results help explaining the mechanisms behind the distribution of barchans in dune fields found on Earth and Mars.

\end{abstract}

\section*{Plain Language Summary}
Barchans are crescent-shaped \corr{wind-formed} dunes \corr{that grow when the wind is roughly unidirectional and the amount of sand is limited (typically, over a non-erodible ground)}. They are found in \corr{dune} fields on Earth and Mars, where \corr{barchan dunes} interact with each other\corr{, forming different field patterns. One first question that arises is why, in some cases, two nearby barchans do not touch each other. Is there a repulsion mechanism?} In this paper, we investigate the barchan-barchan \corr{dune} repulsion by carrying out numerical \corr{computations of the motion of the fluid and each grain forming a pair of barchans. The outputs are detailed measurements of both fluid and grains, such as trajectories, forces, and mass balances, and they show that the downstream barchan shrinks faster than the upstream one. The results also show that the faster shrinkage is due to the fluid flow disturbance caused by the upstream barchan: the fluid impacts the downstream dune, increasing its erosion at the same time that it entrains further downstream part of grains coming from the upstream dune. As a result, the downstream barchan moves at the same pace or faster than the upstream dune, though receiving grains from the latter (which only loses grains).}

\section{Introduction}

Barchans are \corr{eolian} dunes of crescent shape formed by the action of a roughly unidirectional flow over a limited quantity of sand, being commonly found in dune fields on Earth, Mars and other celestial bodies \cite{Bagnold_1, Herrmann_Sauermann, Hersen_3, Elbelrhiti, Claudin_Andreotti, Parteli2, Courrech}. Within those fields, corridors of size-selected barchans are frequently observed, where intricate barchan-barchan interactions have proven essential for size regulation \cite{Hersen_2, Hersen_5, Kocurek, Genois, Genois2, Assis, Assis2, Assis3}. Among the different types of barchan-barchan interaction, there is one, known as chasing \cite{Assis, Assis2}, in which the dunes remain close but without touching each other. This interaction pattern is, in a certain way, counterintuitive, since the downstream dune is in the wake of that upstream. However, \citeA{Assis2} showed that the downstream dune erodes faster than the upstream one, decreasing in size and increasing \corr{in} speed, outrunning then the upstream dune.

The short-range interaction of barchans, including barchan-barchan collision (i.e., when they touch each other), was the object of several studies over the last decades, most of them carrying out measurements of eolian barchans \cite{Norris, Gay, Vermeesch, Elbelrhiti2, Hugenholtz}, experiments with aquatic barchans \cite{Endo2, Hersen_5}, and numerical simulations \cite{Lima, Partelli6, Schwammle2, Duran2, Duran3, Katsuki, Genois2, Zhou2}. Although they increased considerably our knowledge on barchan-barchan interactions, those studies have some drawbacks. In the eolian case, time series for barchan-barchan interactions are incomplete, given their long timescale (of the order of decades). In the case of numerical simulations, they consisted of continuous, simplified discrete, or agent-based models, so that they contained simplifications that precluded some interaction patterns of taking place. In common, these studies measured or computed the dynamics of barchans at the bedform scale, and, therefore, the dynamics at the grain scale was not known.

To the authors' knowledge, the first study to inquire specifically into the dune-dune repulsion mechanism was \citeA{Bacik}, who investigated experimentally the dynamics of a pair of two-dimensional (2D) dunes in a narrow Couette-type circular channel. In their experiments, they placed two piles of grains inside the channel filled with water, and paddles on the water free surface imposed a turbulent flow that deformed the piles into two-dimensional dunes that interacted with each other over long times. For this 2D case, the authors found that turbulent structures formed in the wake of bedforms induce dune-dune repulsion, preventing dunes from touching each other. \citeA{Bacik} inferred that the same mechanism could be also valid for two barchans of comparable size, which was later proved true by \citeA{Assis, Assis2}. As previous studies, \citeA{Bacik} conducted measurements at the bedform scale.

Recently, we inquired into barchan-barchan interactions by conducting experiments in a water \corr{channel} \cite{Assis, Assis2, Assis3}, with measurements carried out at both the bedform and grain scales. In the experiments, the initial configurations (aligned or off-centered), initial conditions, grain types (diameter, density and roundness), pile masses, initial distances, and water flow rates were varied, and from the results we found five interaction patterns for both aligned and off-centered configurations: (i) chasing, when dunes do not touch each other; (ii) merging, when collision occurs and the dunes merge; (iii) exchange, when collision occurs and, just afterward, a small barchan is ejected; (iv) fragmentation-chasing, when collision does not occur and the downstream dune splits; and (v) fragmentation-exchange, when fragmentation initiates before collision with one of the split parts takes place. Although our findings explained some aspects of interactions at the grain scale, information such as forces acting on each grain and the dynamics of hidden (totally or partially buried) grains were not accessible. In another front, following \citeA{Alvarez5, Alvarez7}, we carried out CFD-DEM (computational fluid dynamics - discrete element method) of isolated barchan dunes \cite{Lima2}. In the simulations, we used LES (large eddy simulation) for the fluid, with the smaller scales of the order of the grain diameter. With those simulations, we could measure information missing in experiments, such as those listed above (forces on each grain, for instance). To the best of our knowledge, these are the only grain scale simulations of barchans carried out to this date. If successfully conducted for barchan-barchan interactions, all the missing details for understanding the different patterns would be available.

\corr{More recently, \citeA{He} investigated specifically the chasing pattern of two interacting barchans. For that, they carried out experiments in a water tunnel where two piles of grains that were aligned in the flow direction evolved into barchan dunes that migrated downstream without touching each other. They showed that, although dunes have initially different celerities, they eventually reach approximately the same speed, maintaining their separation distance from that time on. This behavior is similar to that showed by \citeA{Bacik} for 2D dunes in a periodic flume. In order to better explain that behavior, they also carried out CFD simulations in which the dunes consisted in fixed objects (i.e., without the motion of grains) and the fluid was computed using the lattice Boltzmann method. Based on the results, they proposed a model that explains the evolution toward an equilibrium separation, which is based on the difference in migration speeds, caused by large size differences, on a repulsion effect caused by vortical structures of the fluid flow when dunes are very close to each other (hindering dune-dune collision), and on an embracing effect caused also by the vortical structures of the fluid flow, but which decelerates the downstream dune when it is much smaller than and far from the upstream dune. These results shed new light on the mechanisms of barchan-barchan repulsion, but a detailed study at the grain scale is, however, still missing.}

In this paper, we inquire into the origins of the repulsion mechanism of the chasing pattern by carrying out LES-DEM computations. In the simulations, solved at the scale of grains, a pair of granular heaps is deformed by a water flow into barchan dunes that interact with each other, and we compute data such as position, velocity and resultant force for each individual particle at each time step. We show the trajectories of grains, time-average resultant forces, and mass balances for each dune, and that the downstream barchan shrinks faster than the upstream one. Therefore, the downstream barchan keeps a relatively high velocity although in the wake of the upstream one. We also show that, in its turn, the fast shrinkage of the downstream barchan is caused by the flow disturbance, which induces higher erosion on the downstream barchan and its circumvention by grains leaving the upstream dune (so that deposition of grains from the upstream dune is small). Our results represent a contribution for understanding the mechanisms behind the distribution of barchans in dune fields on Earth, Mars, and other celestial bodies.

\section{Materials and Methods}

\begin{sloppypar}
We carried out Euler-Lagrange simulations, in which the fluid flow is computed in an Eulerian grid while the solid particles are tracked in a Lagrangian framework. We used the same model described in \citeA{Lima2}, where the fluid is solved with LES by using the open-source CFD code OpenFOAM (https://openfoam.org), and the motion of grains is solved with DEM by using the open-source code LIGGGHTS \cite{Kloss, Berger}. The coupling between CFD and DEM is done by the open-source code \mbox{CFDEM} \cite<www.cfdem.com,>{Goniva}.
\end{sloppypar}

\corr{In terms of basic equations, the Lagrangian part} computes the linear and angular momentum equations for each solid particle (grain), given by Equations \ref{Fp} and \ref{Tp}, respectively,

\begin{equation}
	m_{p}\frac{d\vec{u}_{p}}{dt}= \vec{F}_{p}\, ,
	\label{Fp}
\end{equation}

\begin{equation}
	I_{p}\frac{d\vec{\omega}_{p}}{dt}=\vec{T}_{c}\, ,
	\label{Tp}
\end{equation}

\noindent where, for each grain, $m_{p}$ is the mass, $\vec{u}_{p}$ is the velocity, $I_{p}$ is the moment of inertia, $\vec{\omega}_{p}$ is the angular velocity, $\vec{T}_{c}$ is the resultant of contact torques between solids, and $\vec{F}_{p}$ is the resultant force (weight, contact and fluid forces), given by

\begin{equation}
	\vec{F}_{p}= \vec{F}_{fp} + \vec{F}_{c} + m_{p}\vec{g}\, ,
	\label{Fp2}
\end{equation}

\noindent In Equation \ref{Fp2}, $\vec{g}$ is the acceleration of gravity and, for each grain, $\vec{F}_{c}$ is the resultant of contact forces between solids and $\vec{F}_{fp}$ is the resultant of fluid forces. For the contact forces and torques, $\vec{F}_{c}$ and $\vec{T}_{c}$, respectively, we consider a Hertzian model, in which contact forces are decomposed into normal and tangential components, as shown briefly in the Supporting Information \cite<for more details, see>{Lima2}. For the resultant of fluid forces, $\vec{F}_{fp}$, we consider Equation \ref{Ffp_sim},

\begin{equation}
	\vec{F}_{fp} = \vec{F}_{d} + \vec{F}_{p} + \vec{F}_{\tau} + \vec{F}_{vm} \, ,
	\label{Ffp_sim}
\end{equation}

\noindent where $\vec{F}_{d}$ is the fluid drag, $\vec{F}_{p}$ is the force due to pressure gradient, $\vec{F}_{\tau}$ is the force due to the deviatoric stress tensor, and $\vec{F}_{vm}$ is the virtual mass force \cite<we neglect the Basset, Saffman, and Magnus forces because they are usually negligible in CFD-DEM simulations,>{Zhou}. As in previous works, we neglect torques caused directly by the fluid in the angular momentum (Equation \ref{Tp}), since those due to contacts are much higher \cite{Tsuji, Tsuji2, Liu}.

Because the fluid is water, the \corr{Eulerian part consists basically of} the incompressible mass and momentum equations, given by Equations \ref{mass} and \ref{mom}, respectively,

\begin{equation}
	\nabla\cdot\vec{u}_{f}=0 \, ,
	\label{mass}
\end{equation}

\begin{equation}
	\frac{\partial{\rho_{f}\vec{u}_{f}}}{\partial{t}} + \nabla \cdot (\rho_{f}\vec{u}_{f}\vec{u}_{f}) = -\nabla P + \nabla\cdot \vec{\vec{\tau}} + \rho_{f}\vec{g} - \frac{N}{V}\vec{F}_{fp} \, ,
	\label{mom}
\end{equation}

\noindent where $\vec{u}_{f}$ is the fluid velocity, $\rho_{f}$ is the fluid density, $P$ the fluid pressure, $\vec{\vec{\tau}}$ the deviatoric stress tensor of the fluid, $\vec{g}$ is the acceleration of gravity, $N$ is the number of grains in a given cell and $V$ is the cell volume. More details of the used model and numerical implementation are available in \citeA{Lima2}.

\begin{figure}[ht]
	\begin{center}
		\includegraphics[width=0.8\linewidth]{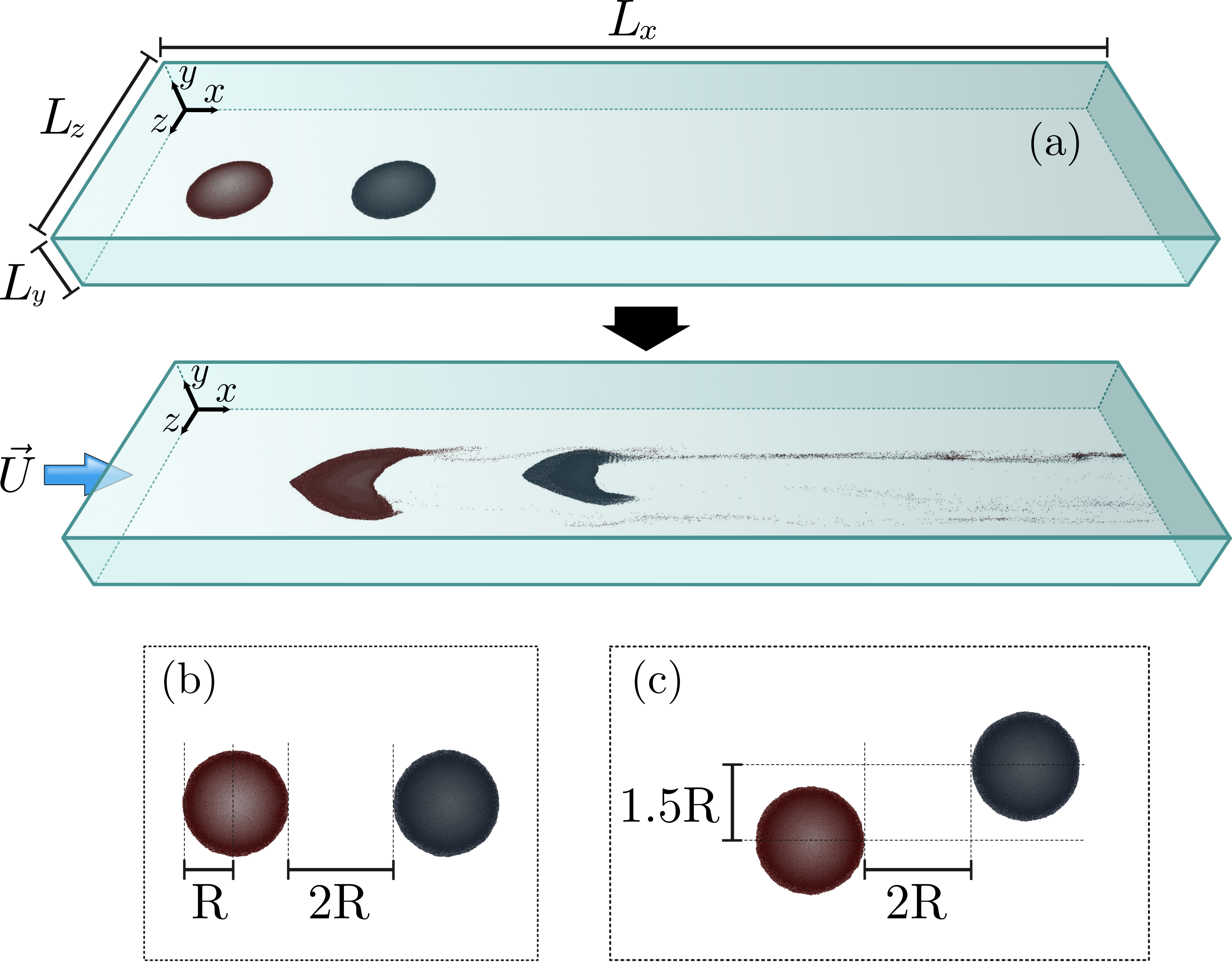}\\
	\end{center}
	\caption{(a) Layout of the numerical setup, showing the channel dimensions, the flow direction, the initial piles, and their evolution at a posterior time. (b) and (c) Relative positions of the initial piles (top view) for the aligned and off-centered cases, respectively. In both cases, the upstream pile was initially placed at 3 cm from the CFD inlet.}
	\label{fig:setup}
\end{figure}

The CFD domain consists in a 3D channel of size $L_x$ = 0.4 m, $L_y$ = $\delta$ = 0.025 m and $L_z$ = 0.1 m, where $x$, $y$ and $z$ are the longitudinal, vertical and spanwise directions, respectively. We note that, for saving computing time, the vertical dimension $L_y$ = $\delta$ corresponds to the channel half height, i.e., the channel centerline (the real channel height is 2$\delta$). With that, the CFD domain has periodic conditions in the longitudinal and spanwise directions, no-slip conditions on the bottom wall, and free slip on the top boundary ($y$ = $\delta$). The channel Reynolds number based on the cross-sectional mean velocity $U$, Re = $U 2\delta \nu^{-1}$, is 14,000, and the Reynolds number based on shear velocity $u_*$, Re$_*$ = $u_* \delta \nu^{-1}$, is 400, where $\nu$ is the kinematic viscosity of the water. The granular material forming each initial heap consisted of 10$^5$ glass spheres, with sizes randomly distributed (in a Gaussian distribution) within 0.15 mm $\leq$ $d$ $\leq$ 0.25 mm\corr{, with mean value of 0.2 mm and standard deviation of 0.025 mm (for reference, the number of particles used in \citeA{Bacik} and \citeA{He} was of the order of 100,000, and within 34,800 and 573,000 in \citeA{Assis}). Under these conditions, the saturation length of bedload $L_{sat}$ is of the order of 1 mm and the most unstable mode of linear instabilities $\lambda_{max}$ of the order of 100 mm \cite{Franklin_12}}. The boundary conditions for the grains were solid wall at the bottom boundary, free exit at the outlet, and no grain influx at the inlet \cite<the number of grains in the domain decreased along time, in the same way as in our previous experiments,>{Assis, Assis2, Assis3}. The current setup is similar to those shown in \citeA{Lima2}, where the DEM and LES parameters were extensively tested and compared with experiments. Therefore, a complete description of CFD meshes and convergence, DEM parameters, and tests and validation can be found in \citeA{Lima2}, and more details of the numerical setup are available in the Supporting Information.

Prior to simulations of barchan-barchan interactions, we carried out LES simulations of pure water flow in the periodic channel, until reaching a fully-developed turbulent flow. The results were stored to be used as initial condition for the fluid in the LES-DEM simulations. The next step was then to completely stop the water flow and let the grains settle by free fall, forming two conical piles with radius $R$ $\approx$ 0.0145 m and height $h$ $\approx$ 0.003 m. The piles were distant 2$R$ from each other in the longitudinal direction, and either 0 (for the aligned case) or 1.5$R$ (for the off-centered case) in transverse direction, as shown in Figures \ref{fig:setup}b and \ref{fig:setup}c, and the upstream pile was initially placed at 3 cm from the CFD inlet. \corr{The size (or mass) ratio of the upstream to the downstream dune was equal to unity, since, as showed by \citeA{Assis}, the chasing pattern occurs when barchans have approximately the same size (of course, for large ratios the upstream barchan will not reach the downstream one if both piles consist of the same type of grains).} The final step was, thus, to impose the turbulent flow stored in a previous step.

\section{Results}

\subsection{Morphology}

\begin{figure}[ht]
	\begin{center}
		\includegraphics[width=0.7\linewidth]{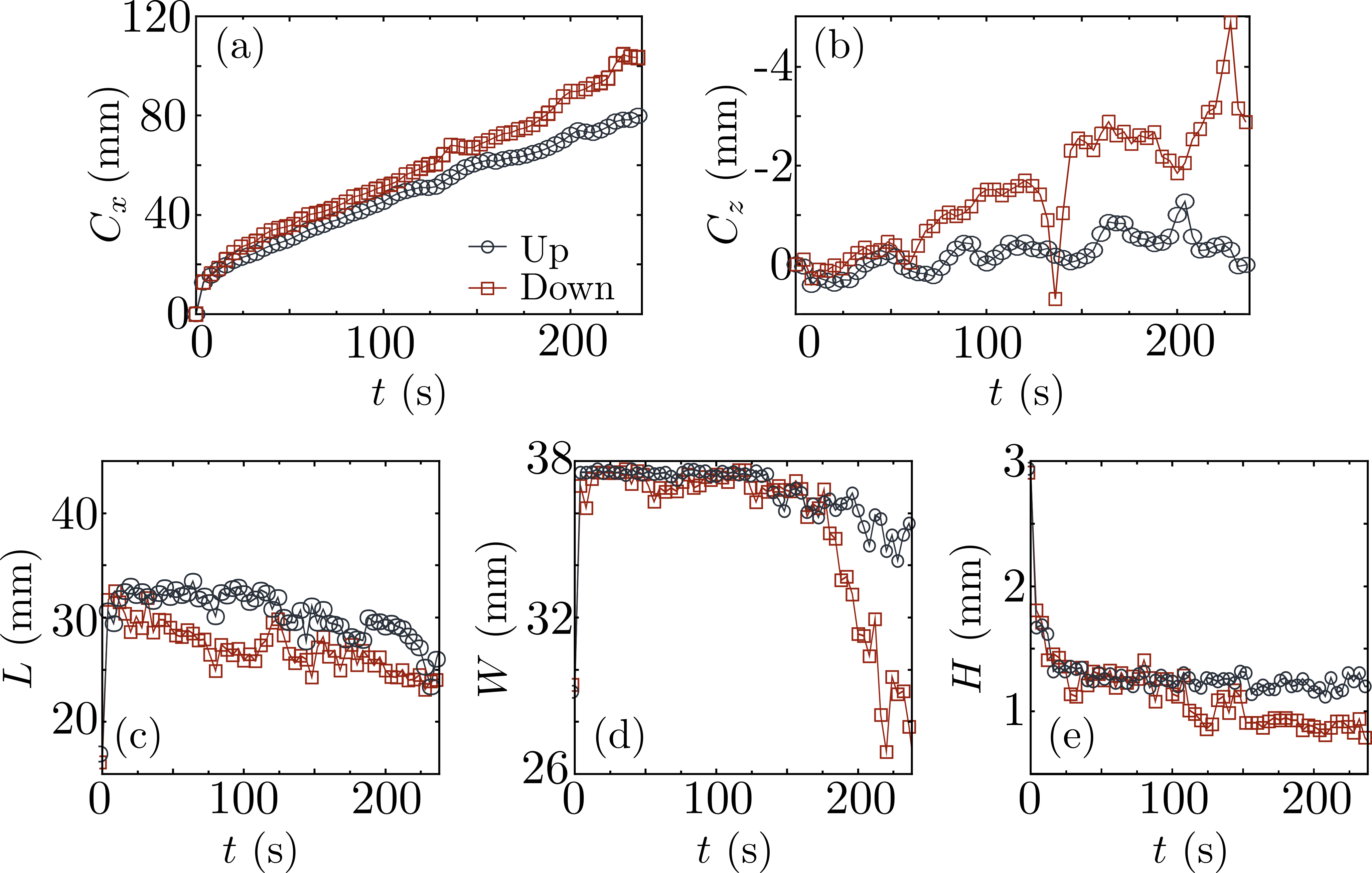}\\
	\end{center}
	\caption{Morphodynamics of interacting barchans in the aligned case. (a) Longitudinal displacement $C_x$, (b) transverse displacement $C_z$, (c) length $L$, (d) width $W$, and (e) height $H$ of barchans. In the graphics, \textit{Up} stands for the upstream barchan and \textit{Down} for the downstream barchan, and we fixed the origins of $C_x$ and $C_z$ of each dune in the centroid of the respective initial pile.}
	\label{fig:morphology_aligned}
\end{figure}

\begin{figure}[ht]
	\begin{center}
		\includegraphics[width=0.7\linewidth]{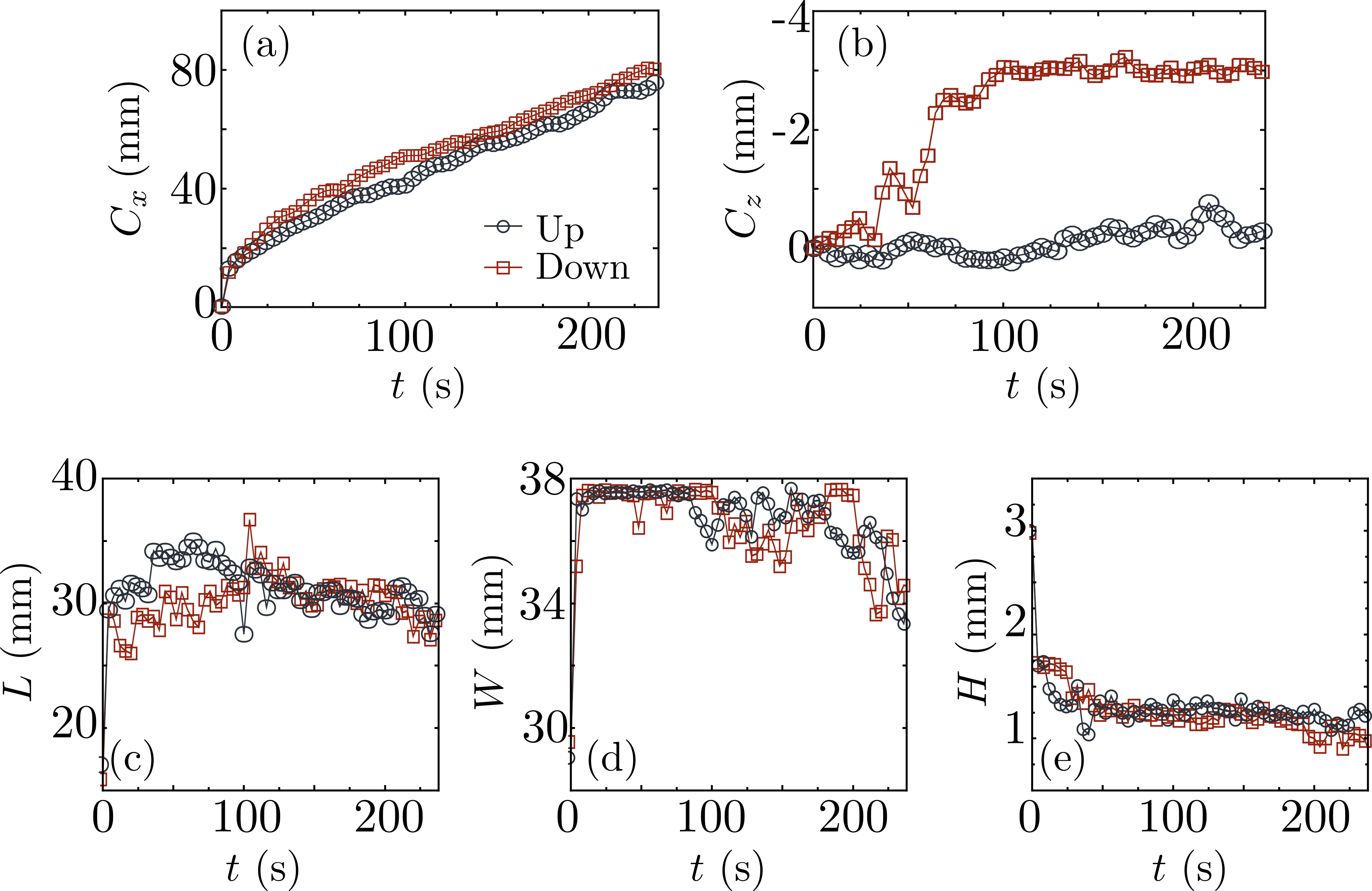}\\
	\end{center}
	\caption{Morphodynamics of interacting barchans in the off-centered case. (a) Longitudinal displacement $C_x$, (b) transverse displacement $C_z$, (c) length $L$, (d) width $W$, and (e) height $H$ of barchans. In the graphics, \textit{Up} stands for the upstream barchan and \textit{Down} for the downstream barchan, and we fixed the origins of $C_x$ and $C_z$ of each dune in the centroid of the respective initial pile.}
	\label{fig:morphology_offcentered}
\end{figure}

As soon as the water flow is imposed in the domain, the conical piles are deformed into two barchan dunes that interact with each other, as can be observed in the snapshots (with some transparency for better observing grain migration) of Figures \ref{fig:grains_align}a and \ref{fig:grains_offcenter}a for the aligned and off-centered cases, respectively. Snapshots without transparency (Figure S1) and movies showing the evolution of bedforms are available in Supporting Information. The behavior for these flow conditions is similar to those observed in our previous experiments \cite{Assis, Assis2}, with the downstream barchan shrinking and moving faster than the upstream one (\textit{chasing} pattern), most noticeable in the aligned case. This is evinced in Figures \ref{fig:morphology_aligned} and \ref{fig:morphology_offcentered}, showing the time evolution of longitudinal and transverse displacements ($C_x$ and $C_z$, panels (a) and (b)), length ($L$, panel (c)), width ($W$, panel (d)), and height ($H$, panel (e)) of dunes for the aligned and off-centered cases, respectively. In Figures \ref{fig:morphology_aligned}a-b and \ref{fig:morphology_offcentered}a-b, the longitudinal and transverse displacements were computed based on the centroid of each dune, and we fixed the origins of $C_x$ and $C_z$ in the centroid of the respective initial pile. In the aligned case, we observe a higher decrease in $L$, $W$ and $H$ for the downstream dune than for the upstream one, while its longitudinal displacement $C_x$ increases faster than that of the upstream barchan, indicating an augmentation of the longitudinal separation along time. In the off-centered case, $L$, $W$, $H$ and $C_x$ vary in a similar way for both barchans, indicating that they keep roughly the same longitudinal  separation along time. The variations in the longitudinal separation are visible in the snapshots shown in Figures \ref{fig:grains_align}a and \ref{fig:grains_offcenter}a.

In both cases, the transverse displacement $C_z$ of the upstream dune remains close to zero, while that of the downstream dune increases in modulus as a consequence of its motion in the transverse direction, as also observed in previous experiments \cite{Assis, Assis2}. The reason for the transverse motion is related to the exchange of grains between barchans, described in Subsection \ref{subsection_exchange}. In the aligned case, we note that Figure \ref{fig:morphology_aligned}b shows two large peaks for $C_z$ due to clumps of grains that are ejected from and/or absorbed by the downstream barchan. \corr{Figure S4 of Supporting Information shows a direct comparison between our numerical results and the experiments.}

\subsection{Motion of grains and mass exchange}
\label{subsection_exchange}

One advantage of discrete simulations in comparison with our previous experiments \cite{Assis, Assis2} is the knowledge of the instantaneous position of all grains, \corr{making it possible to track} each one of them along time \corr{(the velocities and forces of all grains are computed and stored at each time step, and the trajectories are reconstructed by linking the instantaneous positions occupied by the centroid of each considered particle along time)}. With that, typical trajectories and velocities of grains can be computed, as well as the resultant force acting on each of them. Next, we will focus the discussion on grains exchanged between barchans, see Figure S2 in Supporting Information for graphics showing velocity fields of grains moving over barchans.

\begin{figure}[ht]
	\begin{center}
		\includegraphics[width=1\linewidth]{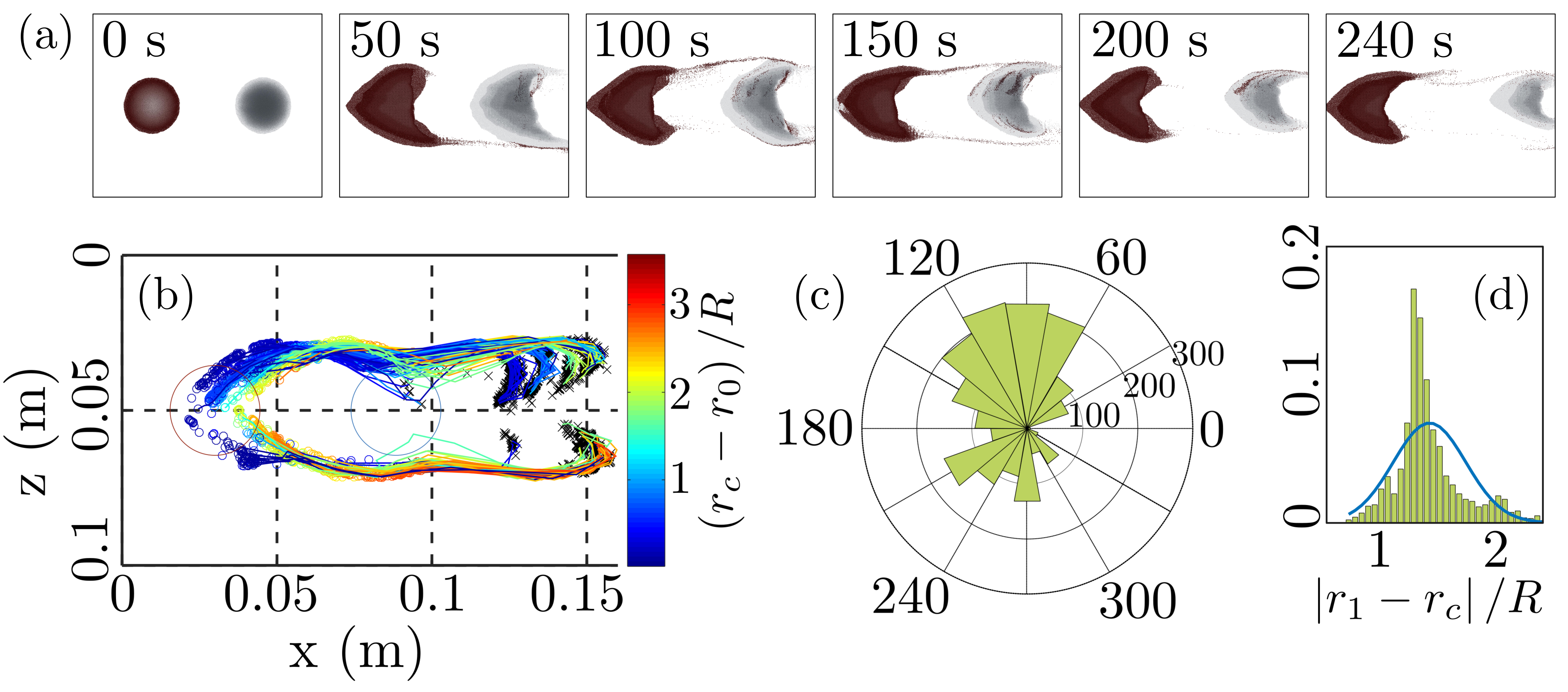}\\
	\end{center}
	\caption{Mass exchange in the aligned case. (a) Snapshots with transparency showing the grains of each dune (top view) at different instants (appearing in maroon for the upstream barchan and gray for the downstream one due to the percentage of transparency adopted). (b) Trajectories of grains leaving the upstream barchan and reaching the downstream one, in which the colorbar indicates the dune longitudinal position when the considered grain started its motion. The large circles indicate the initial piles (top view), the small circles the initial position of the considered grain when it started moving on the upstream barchan, and the x's indicate their respective final positions (when stopping) on the downstream barchan. (c)--(d) Number of grains $N$ from the upstream barchan reaching the downstream one, in polar coordinates (with origin on the centroid of the upstream barchan): (c) frequency of occurrence of $N$ as a function of the angle (the origin is aligned with the flow direction), and (d) probability density function (pdf) of $N$ as a function of the radial position. In the figure, $R$ is the radius of the initial pile, $r_1$ is the initial radial position of grains leaving the upstream barchan, and $r_c$ and $r_0$ are, respectively, the instantaneous and initial positions of the centroid of the upstream dune.}
	\label{fig:grains_align}
\end{figure}

\begin{figure}[ht]
	\begin{center}
		\includegraphics[width=1\linewidth]{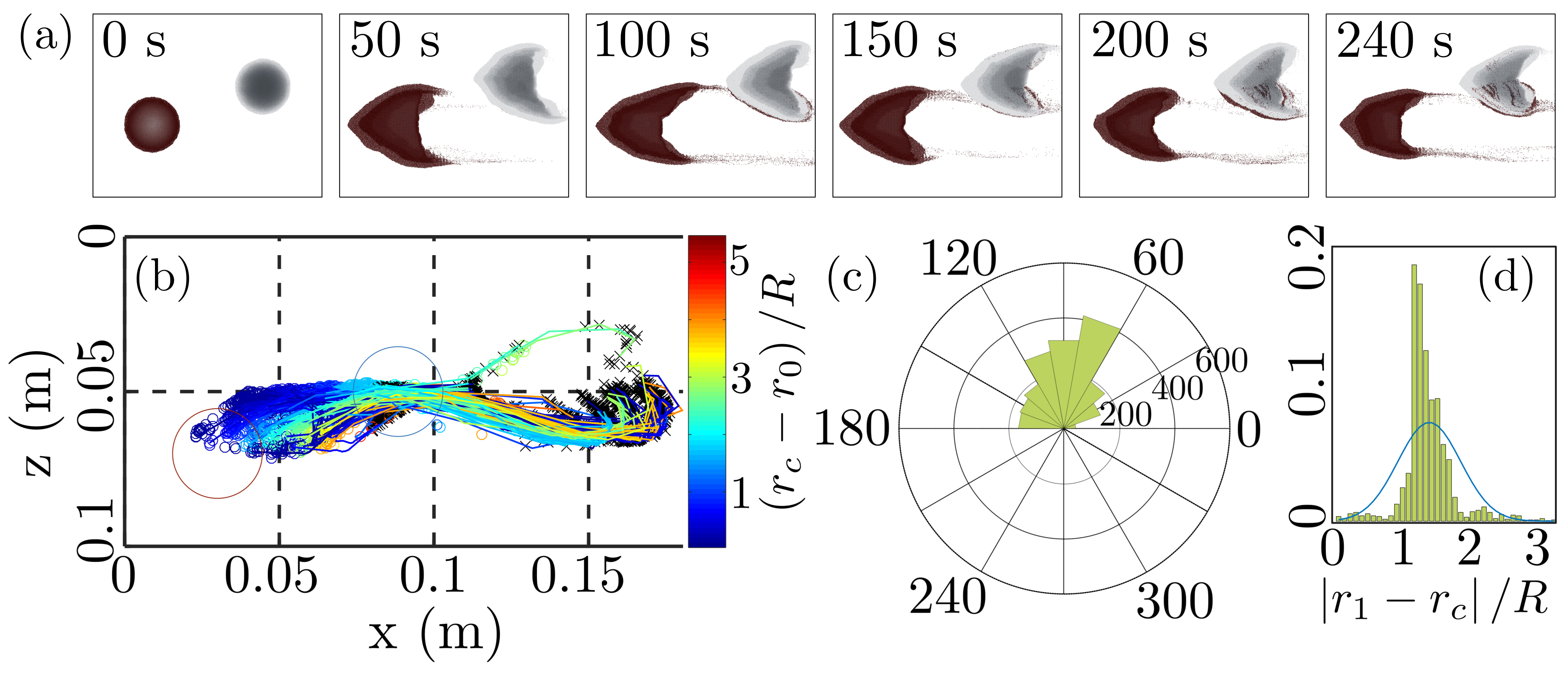}\\
	\end{center}
	\caption{Mass exchange in the off-centered case. (a) Snapshots with transparency showing the grains of each dune (top view) at different instants (appearing in maroon for the upstream barchan and gray for the downstream one due to the percentage of transparency adopted). (b) Trajectories of grains leaving the upstream barchan and reaching the downstream one, in which the colorbar indicates the dune longitudinal position when the considered grain started its motion. The large circles indicate the initial piles (top view), the small circles the initial position of the considered grain when it started moving on the upstream barchan, and the x's indicate their respective final positions (when stopping) on the downstream barchan. (c)--(d) Number of grains $N$ from the upstream barchan reaching the downstream one, in polar coordinates (with origin on the centroid of the upstream barchan): (c) frequency of occurrence of $N$ as a function of the angle (the origin is aligned with the flow direction), and (d) pdf of $N$ as a function of the radial position. In the figure, $R$ is the radius of the initial pile, $r_1$ is the initial radial position of grains leaving the upstream barchan, and $r_c$ and $r_0$ are, respectively, the instantaneous and initial positions of the centroid of the upstream dune.}
	\label{fig:grains_offcenter}
\end{figure}

We, thus, inquired into the trajectories of grains that migrate from one barchan to the other, and identified the points of origin and destination of those grains in order to compute statistics and the mass flow rate by regions. For example, Figures \ref{fig:grains_align}b--d and \ref{fig:grains_offcenter}b--d show a portion of the mass exchange between barchans for the aligned and off-centered cases, respectively. In Figures \ref{fig:grains_align}b and \ref{fig:grains_offcenter}b, we plot trajectories of grains leaving the upstream barchan and reaching the downstream one. In the figures, the colorbar indicates the dune longitudinal position when the considered grain started its motion, where $r_c$ and $r_0$ are, respectively, the instantaneous and initial positions of the centroid of the upstream dune. Large circles represent the initial piles of radius $R$, small circles the initial position of the considered grain when it started moving on the upstream barchan, and the x's indicate their respective final positions (when stopping) on the downstream barchan. We note that we have not plotted in Figures \ref{fig:grains_align}b and \ref{fig:grains_offcenter}b the lines corresponding to all trajectories (in order to avoid saturating the image with trajectory lines). All the circles and x's are, however, plotted in these figures, and we used a velocity threshold corresponding to 0.1$u_*$ for the starting and ending of motions. We first observe (as well as in Figures \ref{fig:grains_align}a and \ref{fig:grains_offcenter}a) that most of those grains, after leaving the upstream dune by its horns (or one of them in the off-centered case), circumvent the downstream dune until arriving in its lee-side/recirculation region, where they accumulate. The major difference between aligned and off-centered cases is that in the aligned case grains leaving both horns of the upstream barchan circumvent the downstream dune, while in the off-centered case only grains from one horn (the one closer to the downstream dune) circumvent the downstream barchan before settling on the lee side. 

In order to analyze the ensemble of those migrating grains, we computed the number of grains $N$ that left the upstream barchan and  reached the downstream one (over 240 s), and identified their respective positions of origin in polar coordinates (with coordinates' origin on the centroid of the upstream barchan). With these numbers, we computed the frequency of occurrence of $N$ as a function of the angle, and the probability density function (pdf) of $N$ as a function of the radial position, which are shown, respectively, in Figures \ref{fig:grains_align}c and \ref{fig:grains_align}d (for the aligned case) and \ref{fig:grains_offcenter}c and \ref{fig:grains_offcenter}d (for the off-centered case), where $r_1$ is the initial radial position of grains leaving the upstream barchan. We observe a large asymmetry in Figure \ref{fig:grains_offcenter}c, which was already expected since the downstream barchan receives grains from just one of the horns of the upstream barchan, but there is also an asymmetry in the aligned case. In this latter case, due to initial fluctuations in the mass exchange, the downstream dune becomes asymmetrical and migrates in the transverse direction toward the horn that sheds more grains. This was observed in our previous experiments \cite{Assis, Assis2}, and can be also observed in Figure  \ref{fig:grains_align}a. Therefore, in the off-centered case most grains migrating to the downstream barchan have their origin in the region close to one of the horns, while in the aligned case the migrating grains have their origin on the flanks of the upstream barchan, with an asymmetry that increases over time as the downstream dune moves in the transverse direction. As a consequence, in the aligned case those grains start moving in upstream regions near the dune flanks, follow a path along the periphery of the upstream barchan until reaching its horns, and from there are shed toward the downstream barchan, as can be seen in the movies S1 and S2 available in the Supporting Information and in those of \citeA{Assis2}. \corr{We note that we have not analyzed how the direction (left or right) of the symmetry breaking is affected by the tested conditions in the aligned case, since this would require a parametric study and the grain-scale simulations are still computationally expensive (this remains to be investigated further). We note also} that in the aligned case part of the grains on the toe of the downstream barchan migrate toward the lee side of the upstream dune entrained by its recirculation region. Trajectories of grains migrating from the downstream to the upstream barchan are available in Figure S3 of Supporting Information. In addition, we computed the velocity field of moving grains, for which we present the time-averaged fields in Figure S2 of Supporting Information. We can observe larger velocities over the upstream barchan than over the downstream one, indicating that the larger erosion over the downstream dune is due to a larger density of moving grains. Indeed, this is the case for the aligned case at all times, and for the off-centered case at the beginning of interactions, as can be seen in the graphics of the density of moving grains shown in Figures S4 and S5 of Supporting Information.

\begin{figure}[ht]
	\begin{center}
		\includegraphics[width=1\linewidth]{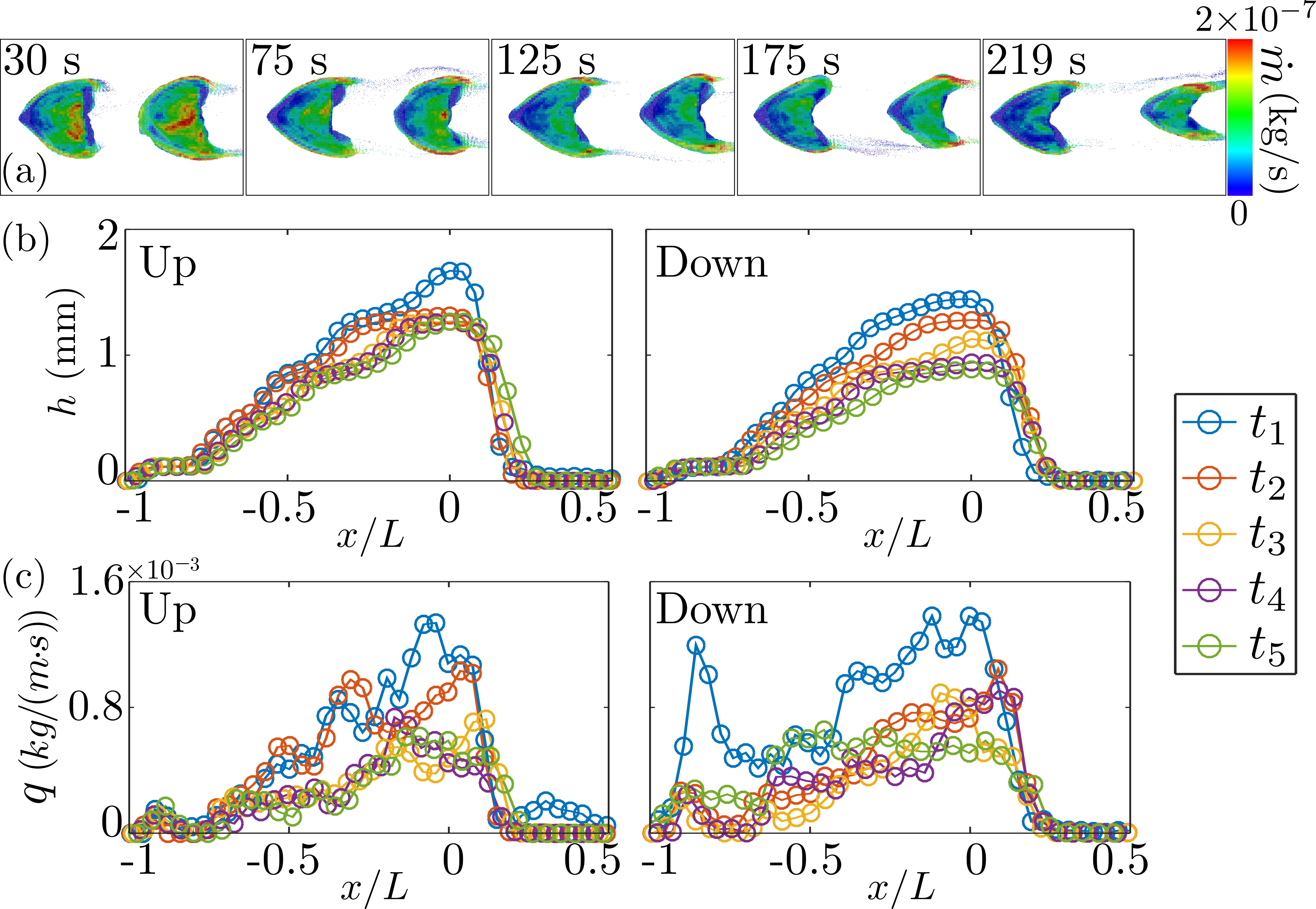}\\
	\end{center}
	\caption{Mass exchange in the aligned case. (a) Snapshots showing the grains of each dune (top view) at different instants, colored in accordance with the average mass flow rate $\dot{m}$ of the region they are in. The averages are computed in intervals $t_1$ to $t_5$ ($t_1$ = 10--50 s, $t_2$ = 51--100 s, $t_3$ = 101--150 s, $t_4$ = 151--200 s, $t_5$ = 201--238 s). (b) Profiles (height $h$ as a function of the dimensionless longitudinal position $x/L$ with origin at the dune crest) of the centerline for the upstream (Up) and downstream (Down) barchans. (c) \corr{Sediment flux $q$ along the barchan (computed based on its central slice only, $\approx$ 1.2 mm thick)}, averaged over the $t_1$ to $t_5$ intervals, for the upstream (Up) and downstream (Down) barchans.}
	\label{fig:mass_exchange_aligned}
\end{figure}

\begin{figure}[ht]
	\begin{center}
		\includegraphics[width=1\linewidth]{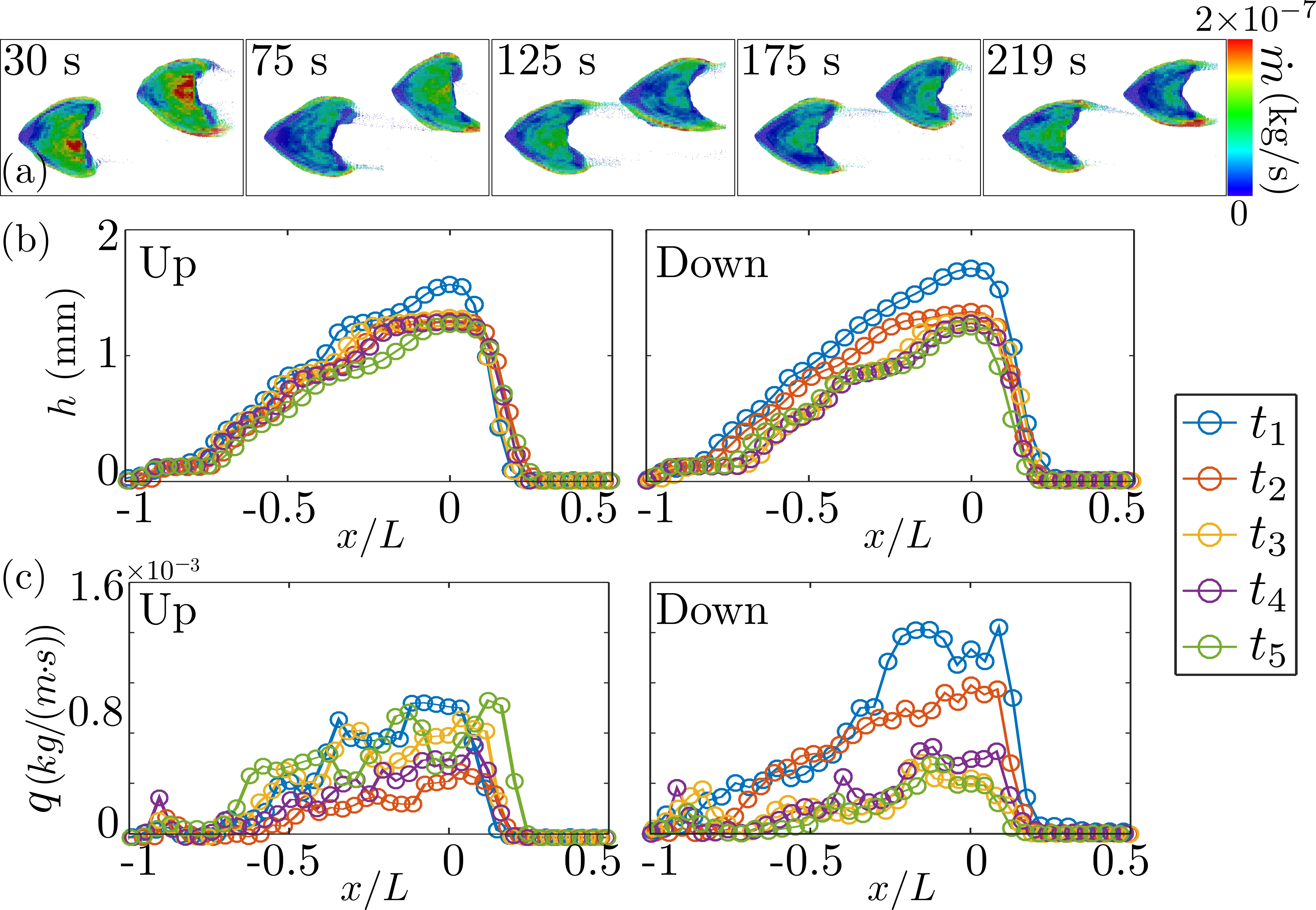}\\
	\end{center}
	\caption{Mass exchange in the off-centered case. (a) Snapshots showing the grains of each dune (top view) at different instants, colored in accordance with the average mass flow rate $\dot{m}$ of the region they are in. The averages are computed in intervals $t_1$ to $t_5$ ($t_1$ = 10--50 s, $t_2$ = 51--100 s, $t_3$ = 101--150 s, $t_4$ = 151--200 s, $t_5$ = 201--238 s). (b) Profiles (height $h$ as a function of the dimensionless longitudinal position $x/L$ with origin at the dune crest) of the centerline for the upstream (Up) and downstream (Down) barchans. (c) \corr{Sediment flux $q$ along the barchan (computed based on its central slice only, $\approx$ 1.2 mm thick)}, averaged over the $t_1$ to $t_5$ intervals, for the upstream (Up) and downstream (Down) barchans.}
	\label{fig:mass_exchange_offcentered}
\end{figure}

In addition to velocities and trajectories, we computed averages of the mass flow rate $\dot{m}$ by counting the number of grains moving over the barchans, which we multiplied by their weight and divided by the corresponding time interval and gravity. The averages are computed in the following time intervals: $t_1$ = 10--50 s, $t_2$ = 51--100 s, $t_3$ = 101--150 s, $t_4$ = 151--200 s, and $t_5$ = 201--238 s. With that, we end with the space distribution of $\dot{m}$ over both barchans, at different stages of the barchan-barchan interaction. For example, Figures \ref{fig:mass_exchange_aligned}a and \ref{fig:mass_exchange_offcentered}a show top views of the grains of each dune for different time instants, colored in accordance with the $\dot{m}$ value of the region they are in, for the aligned and off-centered cases, respectively. We observe that, initially ($t_1$), $\dot{m}$ is higher over the downstream than on the upstream barchan, indicating higher erosion on the former, with a consequent shrinkage and acceleration with respect to the upstream dune. With this data, it is possible to evaluate how $\dot{m}$ varies longitudinally using, for instance, vertical slices cutting the dune. One slice of interest is that passing by the barchan centerline, for which the dune profiles at different intervals are shown in Figures  \ref{fig:mass_exchange_aligned}b and \ref{fig:mass_exchange_offcentered}b for the aligned and off-centered cases, respectively (\textit{Up} referring to the upstream barchan and \textit{Down} to the downstream one). The values of \corr{the sediment flux $q$ = $\dot{m}/W_{slice}$ (where $W_{slice}$ is the width of the considered slice)} along the longitudinal direction are shown in Figures \ref{fig:mass_exchange_aligned}c and \ref{fig:mass_exchange_offcentered}c, for which we computed \corr{$q$} by considering a vertical slice \corr{$W_{slice}$ $\approx$ 1.2 mm thick}. With the exception of the $t_1$ interval \corr{for the aligned case and $t_1$ and $t_2$ for the off-centered case}, we observe approximately the same values of \corr{$q$} for both barchans, while for $t_1$ \corr{in the aligned and $t_1$ and $t_2$ in the off-centered case} the values for the downstream barchan are higher \corr{(for the other time intervals, we prefer to not draw further conclusions given the fluctuations observed in the data)}. Values in the certerline are directly related with the dune celerity, since grains in this region spend long times within the barchan, of the order of many turnover times \cite{Zhang_D}. Therefore, because the downstream dune is the smaller one, similar values of \corr{$q$} in the centerline indicate that it moves faster than the upstream dune, explaining, thus, why the chasing pattern takes place \cite{Assis}. 

For the aligned case, we note a high peak close to the toe of the downstream barchan ($x/L$ $\approx$ -0.75) during the $t_1$ interval, which corresponds to the entrainment of grains from the downstream barchan toward the lee side of the upstream dune (as shown in Figure S3 in Supporting Information). \corr{As a final note, we measured the proportion of grains that left the upstream barchan and ascended the stoss side of the downstream one, and found that they correspond to 3.5 and 8.0\% in the aligned and off-centered cases, respectively.}

\subsection{Resultant force on each grain}
\label{subsection_force}

\begin{figure}[ht]
	\begin{center}
		\includegraphics[width=1\linewidth]{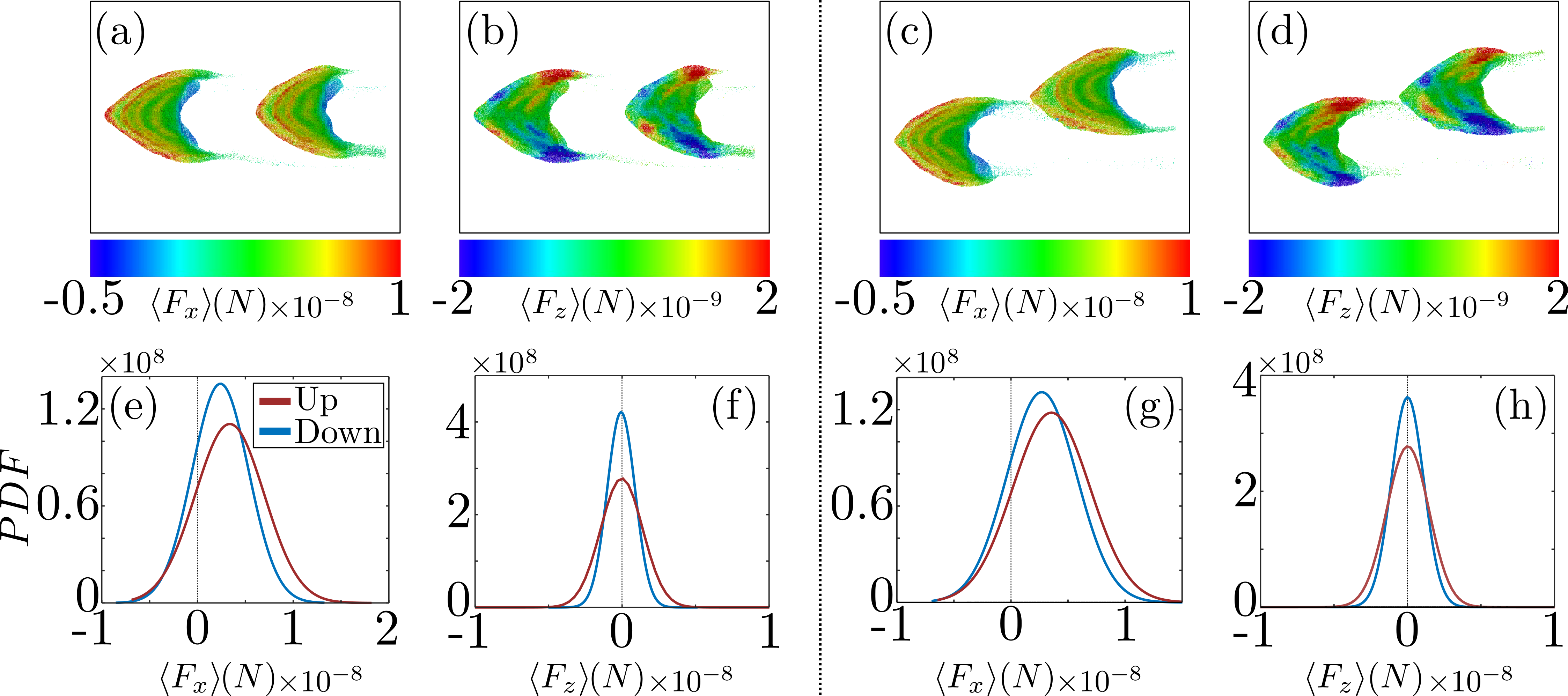}\\
	\end{center}
	\caption{(a)-(d) Top view of dunes colored in accordance with the average resultant force in each region, (a) longitudinal force $\left< F_x \right>$  for aligned dunes; (b) transverse force $\left< F_z \right>$  for aligned dunes; (c) $\left< F_x \right>$ for the off-centered dunes; and (d) $\left< F_z \right>$ for the off-centered dunes. The averages were computed in the 101-150 s interval, by considering all the grains in each region (including those inside the barchans), and the relative position of morphologies plotted in panels (a)-(d) correspond to $t$ = 125 s. The values of forces in N are presented in the colormap below each panel. (e)-(h) Histograms for the of longitudinal $\left< F_x \right>$ and transverse $\left< F_z \right>$ components of forces plotted in the maps of panels (a)-(d), respectively. In the legend, Up and Down stand for upstream and downstream barchans, respectively.}  
	\label{fig:forces_offcenter}
\end{figure}

Because we compute Newton's second law for all particles at each time step, the value of the resultant force acting on each grain is available at all instants. This is a great advantage of discrete computations, since this information is inaccessible from experiments and field measurements. Given the large number of particles in our system (initially 10$^5$ for each dune), we show next plots of the distributions of the resultant force acting on grains, averaged over small time intervals (49 s).

Figures \ref{fig:forces_offcenter}a--d show the distributions of time-averaged resultant forces in different regions of the dune, where Figure \ref{fig:forces_offcenter}a shows the longitudinal force $\left< F_x \right>$ for aligned dunes, Figure \ref{fig:forces_offcenter}b shows the transverse force $\left< F_z \right>$  for aligned dunes, Figure \ref{fig:forces_offcenter}c shows $\left< F_x \right>$ for the off-centered dunes, and Fig. \ref{fig:forces_offcenter}d shows $\left< F_z \right>$ for the off-centered dunes. The averages were computed in the 101-150 s interval, by considering all grains in each region (including those inside the barchans), and the values of forces in N can be read on the colormap below each panel. Figures \ref{fig:forces_offcenter}e--h show histograms for the of longitudinal $\left< F_x \right>$ and transverse $\left< F_z \right>$ components of forces shown in the maps of Figures \ref{fig:forces_offcenter}a--d, respectively. Figures of the instantaneous force acting on specific grains along time (Lagrangian tracking), measured as they move from the upstream barchan toward the downstream one, are shown in \corr{Figure S11} in Supporting Information. We observe from Figures \ref{fig:forces_offcenter}e and \ref{fig:forces_offcenter}g that longitudinal forces on grains of both dunes have distributions that are approximately the same, with the most probable value of downstream dunes having a slightly higher peak at slightly lower values. This shows that, throughout the interaction, grains of both dunes experience similar forces and, therefore, accelerations. This is in agreement with the similar values of $\dot{m}$ found over both dunes (Figures \ref{fig:mass_exchange_aligned} and \ref{fig:mass_exchange_offcentered}). Concerning the transverse component of forces, Figures \ref{fig:forces_offcenter}f and \ref{fig:forces_offcenter}h show that the most probable values are almost the same for both barchans, with grains on the upstream barchan experiencing a slightly larger range of values \corr{(standard deviation 51 and 31\% greater for the upstream barchan with respect to the downstream one, for the aligned and off-centered cases, respectively)}.

\remove{Interestingly, we observe in Figures 8a and 8c some patterns in the form of curved stripes. Those stripes have approximately the same curvature and wavelength as those formed by bidisperse grains over barchans, reported by Alvarez et al. (2021) and Assis et al. (2022). In these works, the authors attributed the stripes to an instability due to grain segregation only, but the distributions of forces might have an important role in their formation as well, although we do not have a physical explanation for the moment.}

\subsection{Fluid flow}

\begin{figure}[ht]
	\begin{center}
		\includegraphics[width=0.6\linewidth]{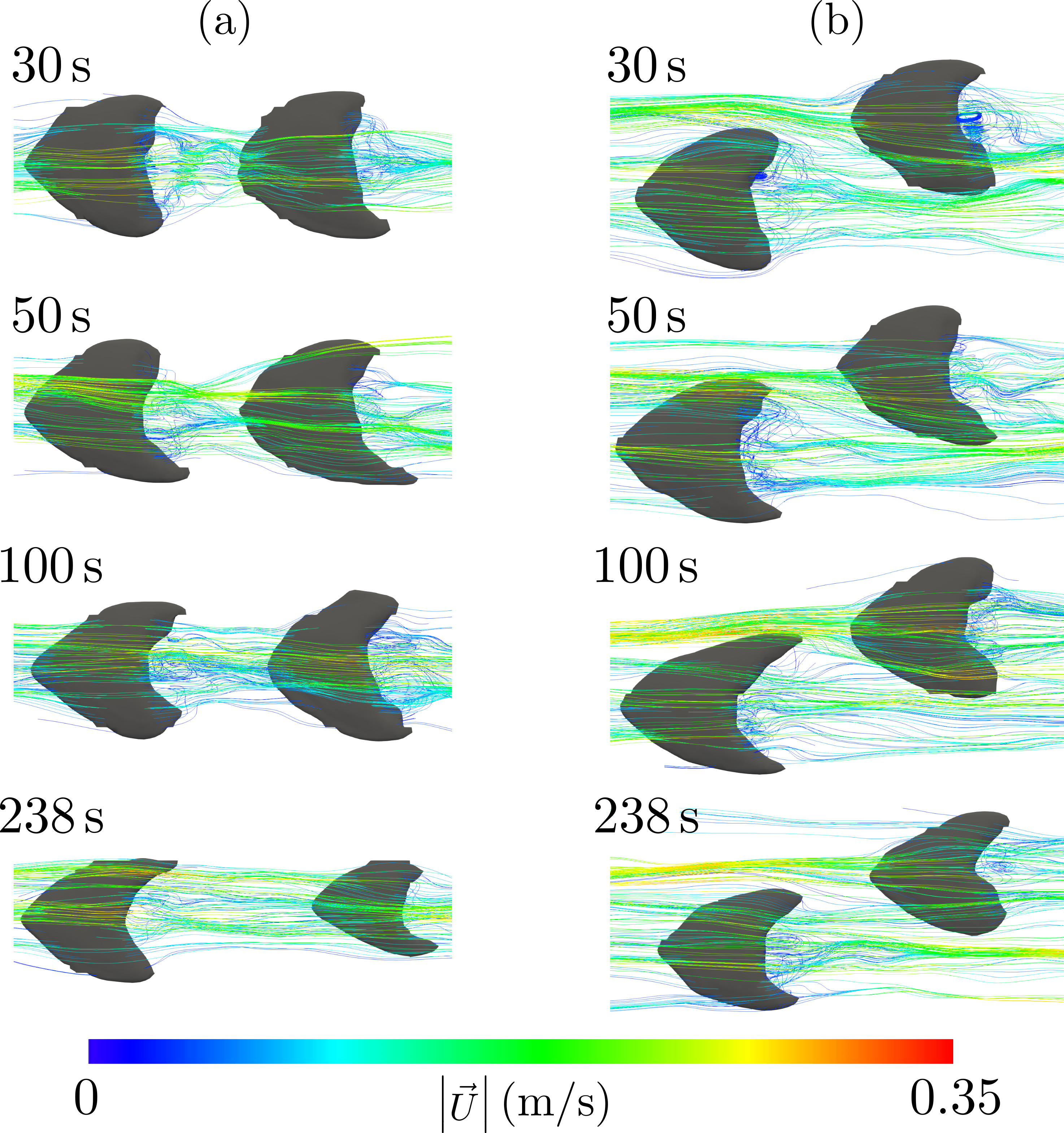}\\
	\end{center}
	\caption{Trajectory lines of the flow over the barchans at different stages of the barchan-barchan interaction for the (a) aligned and (b) off-centered cases. The colors correspond to the magnitude of the velocity of water particles $\left| \vec{U} \right|$, the values of which can be read in the colorbar on the bottom of the figure. The corresponding time instants (within the time interval for the considered trajectories) are shown on the top left of each panel.}
	\label{fig:fluid_flow}
\end{figure}

Finally, we inquire into the fluid flow, which is the mechanism of grain entrainment, and which we computed with a spatial resolution of the order of the grain diameter in the region close to the dune surface. Because of the large quantity of data, we present next typical trajectory lines at different stages of the barchan-barchan interaction.

Figures \ref{fig:fluid_flow}a and \ref{fig:fluid_flow}b show trajectory lines of the water flow over the barchans at different stages of the barchan-barchan interaction, for the aligned and off-centered cases, respectively. These lines shed some light on the behavior of barchans and transport of grains described in previous subsections. For example, for the aligned case, we observe at $t$ = 30 s that the recirculation region of the upstream barchan reaches the downstream dune, explaining why grains from the downstream dune migrate toward the upstream one (as shown in Figure S3 in Supporting information). At later times, we observe that the downstream dune shrinks and moves faster than the upstream dune. In addition, initially small asymmetries in the system make the downstream dune to move in the transverse direction ($t$ = 238 s, for instance). For the off-centered case, we observe that lines from the free-stream flow impact directly half of the downstream dune (in the case shown, from its toe to the exposed horn on the top of the figure). As a consequence, only the trajectory lines that pass over one side of the upstream dune impact the downstream one, which tends to increase or sustain the asymmetry. In addition, in the region between the horn of the upstream barchan and the toe of the downstream one, the upstream flow arrives disturbed by that horn, with a certain amount of acceleration before impacting the downstream barchan \cite<channeling effect, as shown by>{Bristow, Bristow2, Bristow3}, contributing for maintaining the granular mobility and celerity of the downstream barchan. More details on the fluid flow can be seen in Figures S7 to S10 in Supporting Information, where profiles of the mean velocities and second-order moments over both the upstream and downstream barchans are shown.

\corr{\section{Discussion}}

\corr{Our CFD-DEM computations reproduced reasonably well the behavior of the chasing pattern observed in previous experiments, the morphological evolution, migration velocities, and grain trajectories showing good agreement with the experimental results of \citeA{Assis, Assis2}. Besides corroboration of quantities measured experimentally, we obtained the sediment flux over barchans, typical trajectories of the fluid flow, and the resultant force acting on grains. We discuss next the new results of the present work and drawbacks of our simulations.}

\corr{Concerning the time evolution of longitudinal positions of the chasing barchans (Figs. \ref{fig:morphology_aligned}a and \ref{fig:morphology_offcentered}a), we observe that barchans tend to eventually keep a constant separation distance in the off-centered case, while the separation keeps increasing in the aligned case within the total time simulated. On the other hand, the experimental results of \citeA{He} show that in the aligned case barchans eventually reach a constant separation, which contradicts, in a certain way, our results for dune migration \cite<a similar result was shown by>[but for the case of confined 2D dunes]{Bacik}. Therefore, it is possible that simulations of the aligned case carried out over larger times could result in a constant separation distance from a given time instant on. However, different from experiments, grain-scale simulations are expensive to be carried out over long times, which is the case in this particular situation. We nevertheless recognize that simulations over long times would be important for analyzing in detail the separation between barchans, which we suggest as a future work.}

\corr{We found trajectories of grains that agree with those measured by \citeA{Assis2}, but we could analyze a higher number of particles and follow even those that would not be visible in experiments. Besides showing that in the off-centered case grains migrating from the upstream dune to the downstream one have their origin in one of its horns (that closer to the downstream dune) while in the aligned case they have their origin close to the flanks of the upstream dune, and that in the aligned case part of the grains in the toe of the downstream barchan are entrained by the recirculation region toward the upstream dune, we also showed that disturbances of the fluid flow caused by the upstream barchan promote a higher density of moving grains on the downstream dune and, thus, its higher erosion rates. With that, there is a balance between grains received by the downstream barchan and higher erosion rates on its surface, all engendered by disturbances of the fluid flow, which sustains the chasing pattern.}

\corr{We also found the resultant force acting on individual grains, something unfeasible in experiments. We show that the distributions of the longitudinal component of forces are approximately the same for both barchans, so that particles of both dunes undergo similar accelerations, and that the higher erosion rates on the downstream barchan are caused by larger sediment fluxes $q$ at the beginning of interactions. We note here that some patterns in the form of curved stripes can be observed in Figures \ref{fig:forces_offcenter}a and \ref{fig:forces_offcenter}c. Those patters scale with the size of cells used in averaging computations, which leads us to believe that they are of numerical nature.}

\corr{Interestingly, we observe in the movies S1 and S2 of the Supporting Information the formation of a monolayer of grains surrounding each dune. The monolayer is, indeed, observed in the experiments that we carried out in previous works, their width varying between different tests \cite{Alvarez, Alvarez3, Alvarez4, Wenzel, Assis, Assis2}. Our simulations, thus, capture the monolayer, agreeing with experimental observations.}

\corr{Finally, we devoted this paper to the chasing pattern, for which we used simulation conditions in the corresponding space parameters found by \citeA{Assis} (size ratio between barchans close to unity). For this reason, the present simulations did not show any merging or fragmentation behavior, showing a good agreement with the classification map of \citeA{Assis}. We believe that this type of simulation can capture the other patterns observed in experiments with subaqueous barchans, including those with exchange and fragmentation, but this remains to be verified in further studies. We note also that the results presented in this paper are directly applied to subaqueous barchans, but not necessarily to eolian dunes. We understand that the expressive circumvention of grains observed in the subaqueous case is due to the smaller density ratio between grains and fluid ($\rho_p/\rho$ $\approx$ 2.5, where $\rho_p$ is the density of the material of grains), making grains to follow more closely the fluid flow, while in the eolian case the higher density ratio ($\rho_p/\rho$ $\approx$ 2500) engenders saltation (and trajectories aligned in the main flow direction). Therefore, extrapolations of the present results to the eolian case must be done with care. Other issue concerns the total size of the system, which in some cases can make grain-scale simulations unfeasible. In this sense, the use of models working at larger scales, such as the continuous \cite{Sauermann_4,Kroy_C, Schwammle, Parteli4}, agent-based \cite{Genois2, Robson}, or cellular automaton \cite{Narteau, Zhang_D} models, can be useful, when fed with information at the grain scale, for simulating large subaqueous systems.}

\section{Conclusions}

We carried out grain-scale numerical simulations to investigate the mechanisms behind the barchan-barchan \corr{dune} repulsion, an interaction pattern known as \textit{chasing} \cite{Assis}. We showed that, with the exception of the beginning of interactions, where the mass flow rate $\dot{m}$ is greater over the downstream dune, $\dot{m}$ is roughly the same for both barchans, meaning a higher erosion rate on the downstream dune (since it is smaller due to the higher initial $\dot{m}$). We showed also that in the aligned case a great part of the grains leaving the upstream barchan reach the downstream one, but part of grains of the latter migrate to the upstream dune entrained by its recirculation region. In the off-centered case, only grains from one of the horns of the upstream dune reach the downstream one. The transport of grains is corroborated by the trajectory lines of the fluid, which, in the aligned case, show that the recirculation region of the upstream dune reaches the toe of the downstream barchan and, afterward, small asymmetries make more lines from one side of the upstream dune to reach the downstream one, causing the transverse motion of the latter. In the off-centered case, only approximately half of the lines passing over the upstream barchan reach the downstream dune, increasing or sustaining the off-centered configuration. Interestingly, the velocities of particles is slightly higher over the upstream dune, indicating that the higher erosion over the downstream barchan is due to a higher density of moving particles. Finally, we measured the resultant force acting on each grain and showed that the longitudinal component of time-averaged forces are similar for both dunes. Our results shed light on the reasons for the repulsion characteristic of the chasing pattern, helping to explain why sometimes barchans never touch each other. In addition, our findings can be used to refine current large-scale models (such as continuum models), determine the best sites for placing sensors in field studies or carrying out remote sensing, and feed convolutional neural networks (CNNs) for analyzing large datasets.

\section*{Open Research}
\begin{sloppypar}
	Data supporting this work were generated by ourselves and are available in Mendeley Data \cite{Supplemental2} under the CC-BY-4.0 license. The numerical scripts used to post-process the numerical outputs are also available in Mendeley Data \cite{Supplemental2} under the CC-BY-4.0 license.
\end{sloppypar}

\acknowledgments
\begin{sloppypar}
The authors are grateful to FAPESP (Grant Nos. 2018/14981-7, 2019/10239-7 and 2019/20888-2) and to CNPq (Grant No. 405512/2022-8) for the financial support provided.
\end{sloppypar}

\bibliography{references}

\begin{thebibliography}{}

\bibitem [\protect \citeauthoryear {%
Alvarez%
\ \BBA {} Franklin%
}{%
Alvarez%
\ \BBA {} Franklin%
}{%
{\protect \APACyear {2017}}%
}]{%
Alvarez}
\APACinsertmetastar {%
Alvarez}%
\begin{APACrefauthors}%
Alvarez, C\BPBI A.%
\BCBT {}\ \BBA {} Franklin, E\BPBI M.%
\end{APACrefauthors}%
\unskip\
\newblock
\APACrefYearMonthDay{2017}{Dec}{}.
\newblock
{\BBOQ}\APACrefatitle {Birth of a subaqueous barchan dune} {Birth of a
  subaqueous barchan dune}.{\BBCQ}
\newblock
\APACjournalVolNumPages{Phys. Rev. E}{96}{}{062906}.
\newblock
\begin{APACrefURL} \url{https://link.aps.org/doi/10.1103/PhysRevE.96.062906}
  \end{APACrefURL}
\newblock
\begin{APACrefDOI} \doi{10.1103/PhysRevE.96.062906} \end{APACrefDOI}
\PrintBackRefs{\CurrentBib}

\bibitem [\protect \citeauthoryear {%
Alvarez%
\ \BBA {} Franklin%
}{%
Alvarez%
\ \BBA {} Franklin%
}{%
{\protect \APACyear {2018}}%
}]{%
Alvarez3}
\APACinsertmetastar {%
Alvarez3}%
\begin{APACrefauthors}%
Alvarez, C\BPBI A.%
\BCBT {}\ \BBA {} Franklin, E\BPBI M.%
\end{APACrefauthors}%
\unskip\
\newblock
\APACrefYearMonthDay{2018}{Oct}{}.
\newblock
{\BBOQ}\APACrefatitle {Role of Transverse Displacements in the Formation of
  Subaqueous Barchan Dunes} {Role of transverse displacements in the formation
  of subaqueous barchan dunes}.{\BBCQ}
\newblock
\APACjournalVolNumPages{Phys. Rev. Lett.}{121}{}{164503}.
\newblock
\begin{APACrefURL}
  \url{https://link.aps.org/doi/10.1103/PhysRevLett.121.164503}
  \end{APACrefURL}
\newblock
\begin{APACrefDOI} \doi{10.1103/PhysRevLett.121.164503} \end{APACrefDOI}
\PrintBackRefs{\CurrentBib}

\bibitem [\protect \citeauthoryear {%
Alvarez%
\ \BBA {} Franklin%
}{%
Alvarez%
\ \BBA {} Franklin%
}{%
{\protect \APACyear {2019}}%
}]{%
Alvarez4}
\APACinsertmetastar {%
Alvarez4}%
\begin{APACrefauthors}%
Alvarez, C\BPBI A.%
\BCBT {}\ \BBA {} Franklin, E\BPBI M.%
\end{APACrefauthors}%
\unskip\
\newblock
\APACrefYearMonthDay{2019}{Oct}{}.
\newblock
{\BBOQ}\APACrefatitle {Horns of subaqueous barchan dunes: A study at the grain
  scale} {Horns of subaqueous barchan dunes: A study at the grain
  scale}.{\BBCQ}
\newblock
\APACjournalVolNumPages{Phys. Rev. E}{100}{}{042904}.
\newblock
\begin{APACrefURL} \url{https://link.aps.org/doi/10.1103/PhysRevE.100.042904}
  \end{APACrefURL}
\newblock
\begin{APACrefDOI} \doi{10.1103/PhysRevE.100.042904} \end{APACrefDOI}
\PrintBackRefs{\CurrentBib}

\bibitem [\protect \citeauthoryear {%
Alvarez%
\ \BBA {} Franklin%
}{%
Alvarez%
\ \BBA {} Franklin%
}{%
{\protect \APACyear {2020}}%
}]{%
Alvarez5}
\APACinsertmetastar {%
Alvarez5}%
\begin{APACrefauthors}%
Alvarez, C\BPBI A.%
\BCBT {}\ \BBA {} Franklin, E\BPBI M.%
\end{APACrefauthors}%
\unskip\
\newblock
\APACrefYearMonthDay{2020}{Jan}{}.
\newblock
{\BBOQ}\APACrefatitle {Shape evolution of numerically obtained subaqueous
  barchan dunes} {Shape evolution of numerically obtained subaqueous barchan
  dunes}.{\BBCQ}
\newblock
\APACjournalVolNumPages{Phys. Rev. E}{101}{}{012905}.
\newblock
\begin{APACrefURL} \url{https://link.aps.org/doi/10.1103/PhysRevE.101.012905}
  \end{APACrefURL}
\newblock
\begin{APACrefDOI} \doi{10.1103/PhysRevE.101.012905} \end{APACrefDOI}
\PrintBackRefs{\CurrentBib}

\bibitem [\protect \citeauthoryear {%
Alvarez%
\ \BBA {} Franklin%
}{%
Alvarez%
\ \BBA {} Franklin%
}{%
{\protect \APACyear {2021}}%
}]{%
Alvarez7}
\APACinsertmetastar {%
Alvarez7}%
\begin{APACrefauthors}%
Alvarez, C\BPBI A.%
\BCBT {}\ \BBA {} Franklin, E\BPBI M.%
\end{APACrefauthors}%
\unskip\
\newblock
\APACrefYearMonthDay{2021}{}{}.
\newblock
{\BBOQ}\APACrefatitle {Force distribution within a barchan dune} {Force
  distribution within a barchan dune}.{\BBCQ}
\newblock
\APACjournalVolNumPages{Phys. Fluids}{33}{1}{013313}.
\PrintBackRefs{\CurrentBib}

\bibitem [\protect \citeauthoryear {%
Assis%
, Cúñez%
\BCBL {}\ \BBA {} Franklin%
}{%
Assis%
\ \protect \BOthers {.}}{%
{\protect \APACyear {2022}}%
}]{%
Assis3}
\APACinsertmetastar {%
Assis3}%
\begin{APACrefauthors}%
Assis, W\BPBI R.%
, Cúñez, F\BPBI D.%
\BCBL {}\ \BBA {} Franklin, E\BPBI M.%
\end{APACrefauthors}%
\unskip\
\newblock
\APACrefYearMonthDay{2022}{}{}.
\newblock
{\BBOQ}\APACrefatitle {Revealing the Intricate Dune-Dune Interactions of
  Bidisperse Barchans} {Revealing the intricate dune-dune interactions of
  bidisperse barchans}.{\BBCQ}
\newblock
\APACjournalVolNumPages{J. Geophys. Res.: Earth Surf.}{127}{5}{e2021JF006588}.
\newblock
\begin{APACrefURL}
  \url{https://agupubs.onlinelibrary.wiley.com/doi/abs/10.1029/2021JF006588}
  \end{APACrefURL}
\newblock
\begin{APACrefDOI} \doi{https://doi.org/10.1029/2021JF006588} \end{APACrefDOI}
\PrintBackRefs{\CurrentBib}

\bibitem [\protect \citeauthoryear {%
Assis%
\ \BBA {} Franklin%
}{%
Assis%
\ \BBA {} Franklin%
}{%
{\protect \APACyear {2020}}%
}]{%
Assis}
\APACinsertmetastar {%
Assis}%
\begin{APACrefauthors}%
Assis, W\BPBI R.%
\BCBT {}\ \BBA {} Franklin, E\BPBI M.%
\end{APACrefauthors}%
\unskip\
\newblock
\APACrefYearMonthDay{2020}{}{}.
\newblock
{\BBOQ}\APACrefatitle {A Comprehensive Picture for Binary Interactions of
  Subaqueous Barchans} {A comprehensive picture for binary interactions of
  subaqueous barchans}.{\BBCQ}
\newblock
\APACjournalVolNumPages{Geophys. Res. Lett.}{47}{18}{e2020GL089464}.
\PrintBackRefs{\CurrentBib}

\bibitem [\protect \citeauthoryear {%
Assis%
\ \BBA {} Franklin%
}{%
Assis%
\ \BBA {} Franklin%
}{%
{\protect \APACyear {2021}}%
}]{%
Assis2}
\APACinsertmetastar {%
Assis2}%
\begin{APACrefauthors}%
Assis, W\BPBI R.%
\BCBT {}\ \BBA {} Franklin, E\BPBI M.%
\end{APACrefauthors}%
\unskip\
\newblock
\APACrefYearMonthDay{2021}{}{}.
\newblock
{\BBOQ}\APACrefatitle {Morphodynamics of Barchan-Barchan Interactions
  Investigated at the Grain Scale} {Morphodynamics of barchan-barchan
  interactions investigated at the grain scale}.{\BBCQ}
\newblock
\APACjournalVolNumPages{J. Geophys. Res.: Earth Surf.}{126}{8}{e2021JF006237}.
\PrintBackRefs{\CurrentBib}

\bibitem [\protect \citeauthoryear {%
Bacik%
, Lovett%
, Caulfield%
\BCBL {}\ \BBA {} Vriend%
}{%
Bacik%
\ \protect \BOthers {.}}{%
{\protect \APACyear {2020}}%
}]{%
Bacik}
\APACinsertmetastar {%
Bacik}%
\begin{APACrefauthors}%
Bacik, K\BPBI A.%
, Lovett, S.%
, Caulfield, C\BHBI c\BPBI P.%
\BCBL {}\ \BBA {} Vriend, N\BPBI M.%
\end{APACrefauthors}%
\unskip\
\newblock
\APACrefYearMonthDay{2020}{Feb}{}.
\newblock
{\BBOQ}\APACrefatitle {Wake Induced Long Range Repulsion of Aqueous Dunes}
  {Wake induced long range repulsion of aqueous dunes}.{\BBCQ}
\newblock
\APACjournalVolNumPages{Phys. Rev. Lett.}{124}{}{054501}.
\newblock
\begin{APACrefURL}
  \url{https://link.aps.org/doi/10.1103/PhysRevLett.124.054501}
  \end{APACrefURL}
\newblock
\begin{APACrefDOI} \doi{10.1103/PhysRevLett.124.054501} \end{APACrefDOI}
\PrintBackRefs{\CurrentBib}

\bibitem [\protect \citeauthoryear {%
Bagnold%
}{%
Bagnold%
}{%
{\protect \APACyear {1941}}%
}]{%
Bagnold_1}
\APACinsertmetastar {%
Bagnold_1}%
\begin{APACrefauthors}%
Bagnold, R\BPBI A.%
\end{APACrefauthors}%
\unskip\
\newblock
\APACrefYear{1941}.
\newblock
\APACrefbtitle {The Physics of Blown Sand and Desert Dunes} {The physics of
  blown sand and desert dunes}.
\newblock
\APACaddressPublisher{London}{Chapman and Hall}.
\PrintBackRefs{\CurrentBib}

\bibitem [\protect \citeauthoryear {%
Berger%
, Kloss%
, Kohlmeyer%
\BCBL {}\ \BBA {} Pirker%
}{%
Berger%
\ \protect \BOthers {.}}{%
{\protect \APACyear {2015}}%
}]{%
Berger}
\APACinsertmetastar {%
Berger}%
\begin{APACrefauthors}%
Berger, R.%
, Kloss, C.%
, Kohlmeyer, A.%
\BCBL {}\ \BBA {} Pirker, S.%
\end{APACrefauthors}%
\unskip\
\newblock
\APACrefYearMonthDay{2015}{}{}.
\newblock
{\BBOQ}\APACrefatitle {Hybrid parallelization of the {LIGGGHTS} open-source
  {DEM} code} {Hybrid parallelization of the {LIGGGHTS} open-source {DEM}
  code}.{\BBCQ}
\newblock
\APACjournalVolNumPages{Powder Technology}{278}{}{234-247}.
\PrintBackRefs{\CurrentBib}

\bibitem [\protect \citeauthoryear {%
Bristow%
, Blois%
, Best%
\BCBL {}\ \BBA {} Christensen%
}{%
Bristow%
\ \protect \BOthers {.}}{%
{\protect \APACyear {2018}}%
}]{%
Bristow}
\APACinsertmetastar {%
Bristow}%
\begin{APACrefauthors}%
Bristow, N\BPBI R.%
, Blois, G.%
, Best, J\BPBI L.%
\BCBL {}\ \BBA {} Christensen, K\BPBI T.%
\end{APACrefauthors}%
\unskip\
\newblock
\APACrefYearMonthDay{2018}{}{}.
\newblock
{\BBOQ}\APACrefatitle {Turbulent Flow Structure Associated With Collision
  Between Laterally Offset, Fixed-Bed Barchan Dunes} {Turbulent flow structure
  associated with collision between laterally offset, fixed-bed barchan
  dunes}.{\BBCQ}
\newblock
\APACjournalVolNumPages{J. Geophys. Res.-Earth}{123}{9}{2157-2188}.
\PrintBackRefs{\CurrentBib}

\bibitem [\protect \citeauthoryear {%
Bristow%
, Blois%
, Best%
\BCBL {}\ \BBA {} Christensen%
}{%
Bristow%
\ \protect \BOthers {.}}{%
{\protect \APACyear {2019}}%
}]{%
Bristow2}
\APACinsertmetastar {%
Bristow2}%
\begin{APACrefauthors}%
Bristow, N\BPBI R.%
, Blois, G.%
, Best, J\BPBI L.%
\BCBL {}\ \BBA {} Christensen, K\BPBI T.%
\end{APACrefauthors}%
\unskip\
\newblock
\APACrefYearMonthDay{2019}{}{}.
\newblock
{\BBOQ}\APACrefatitle {Spatial Scales of Turbulent Flow Structures Associated
  With Interacting Barchan Dunes} {Spatial scales of turbulent flow structures
  associated with interacting barchan dunes}.{\BBCQ}
\newblock
\APACjournalVolNumPages{J. Geophys. Res.-Earth}{124}{5}{1175-1200}.
\PrintBackRefs{\CurrentBib}

\bibitem [\protect \citeauthoryear {%
Bristow%
, Blois%
, Best%
\BCBL {}\ \BBA {} Christensen%
}{%
Bristow%
\ \protect \BOthers {.}}{%
{\protect \APACyear {2020}}%
}]{%
Bristow3}
\APACinsertmetastar {%
Bristow3}%
\begin{APACrefauthors}%
Bristow, N\BPBI R.%
, Blois, G.%
, Best, J\BPBI L.%
\BCBL {}\ \BBA {} Christensen, K\BPBI T.%
\end{APACrefauthors}%
\unskip\
\newblock
\APACrefYearMonthDay{2020}{}{}.
\newblock
{\BBOQ}\APACrefatitle {Secondary Flows and Vortex Structure Associated With
  Isolated and Interacting Barchan Dunes} {Secondary flows and vortex structure
  associated with isolated and interacting barchan dunes}.{\BBCQ}
\newblock
\APACjournalVolNumPages{J. Geophys. Res.-Earth}{125}{2}{e2019JF005257}.
\PrintBackRefs{\CurrentBib}

\bibitem [\protect \citeauthoryear {%
Claudin%
\ \BBA {} Andreotti%
}{%
Claudin%
\ \BBA {} Andreotti%
}{%
{\protect \APACyear {2006}}%
}]{%
Claudin_Andreotti}
\APACinsertmetastar {%
Claudin_Andreotti}%
\begin{APACrefauthors}%
Claudin, P.%
\BCBT {}\ \BBA {} Andreotti, B.%
\end{APACrefauthors}%
\unskip\
\newblock
\APACrefYearMonthDay{2006}{}{}.
\newblock
{\BBOQ}\APACrefatitle {A scaling law for aeolian dunes on {M}ars, {V}enus,
  {E}arth, and for subaqueous ripples} {A scaling law for aeolian dunes on
  {M}ars, {V}enus, {E}arth, and for subaqueous ripples}.{\BBCQ}
\newblock
\APACjournalVolNumPages{Earth Plan. Sci. Lett.}{252}{}{20-44}.
\PrintBackRefs{\CurrentBib}

\bibitem [\protect \citeauthoryear {%
{Courrech du Pont}%
}{%
{Courrech du Pont}%
}{%
{\protect \APACyear {2015}}%
}]{%
Courrech}
\APACinsertmetastar {%
Courrech}%
\begin{APACrefauthors}%
{Courrech du Pont}, S.%
\end{APACrefauthors}%
\unskip\
\newblock
\APACrefYearMonthDay{2015}{}{}.
\newblock
{\BBOQ}\APACrefatitle {Dune morphodynamics} {Dune morphodynamics}.{\BBCQ}
\newblock
\APACjournalVolNumPages{C. R. Phys.}{16}{1}{118 - 138}.
\PrintBackRefs{\CurrentBib}

\bibitem [\protect \citeauthoryear {%
Dur\'an%
, Schw\"ammle%
\BCBL {}\ \BBA {} Herrmann%
}{%
Dur\'an%
\ \protect \BOthers {.}}{%
{\protect \APACyear {2005}}%
}]{%
Duran2}
\APACinsertmetastar {%
Duran2}%
\begin{APACrefauthors}%
Dur\'an, O.%
, Schw\"ammle, V.%
\BCBL {}\ \BBA {} Herrmann, H.%
\end{APACrefauthors}%
\unskip\
\newblock
\APACrefYearMonthDay{2005}{Aug}{}.
\newblock
{\BBOQ}\APACrefatitle {Breeding and solitary wave behavior of dunes} {Breeding
  and solitary wave behavior of dunes}.{\BBCQ}
\newblock
\APACjournalVolNumPages{Phys. Rev. E}{72}{}{021308}.
\newblock
\begin{APACrefURL} \url{https://link.aps.org/doi/10.1103/PhysRevE.72.021308}
  \end{APACrefURL}
\newblock
\begin{APACrefDOI} \doi{10.1103/PhysRevE.72.021308} \end{APACrefDOI}
\PrintBackRefs{\CurrentBib}

\bibitem [\protect \citeauthoryear {%
Dur\'an%
, Schw\"ammle%
, Lind%
\BCBL {}\ \BBA {} Herrmann%
}{%
Dur\'an%
\ \protect \BOthers {.}}{%
{\protect \APACyear {2009}}%
}]{%
Duran3}
\APACinsertmetastar {%
Duran3}%
\begin{APACrefauthors}%
Dur\'an, O.%
, Schw\"ammle, V.%
, Lind, P\BPBI G.%
\BCBL {}\ \BBA {} Herrmann, H.%
\end{APACrefauthors}%
\unskip\
\newblock
\APACrefYearMonthDay{2009}{}{}.
\newblock
{\BBOQ}\APACrefatitle {The dune size distribution and scaling relations of
  barchan dune fields} {The dune size distribution and scaling relations of
  barchan dune fields}.{\BBCQ}
\newblock
\APACjournalVolNumPages{Granular Matter}{11}{}{7-11}.
\PrintBackRefs{\CurrentBib}

\bibitem [\protect \citeauthoryear {%
Elbelrhiti%
, Andreotti%
\BCBL {}\ \BBA {} Claudin%
}{%
Elbelrhiti%
\ \protect \BOthers {.}}{%
{\protect \APACyear {2008}}%
}]{%
Elbelrhiti2}
\APACinsertmetastar {%
Elbelrhiti2}%
\begin{APACrefauthors}%
Elbelrhiti, H.%
, Andreotti, B.%
\BCBL {}\ \BBA {} Claudin, P.%
\end{APACrefauthors}%
\unskip\
\newblock
\APACrefYearMonthDay{2008}{}{}.
\newblock
{\BBOQ}\APACrefatitle {Barchan dune corridors: Field characterization and
  investigation of control parameters} {Barchan dune corridors: Field
  characterization and investigation of control parameters}.{\BBCQ}
\newblock
\APACjournalVolNumPages{J. Geophys. Res.: Earth Surf.}{113}{F2}{}.
\PrintBackRefs{\CurrentBib}

\bibitem [\protect \citeauthoryear {%
Elbelrhiti%
, Claudin%
\BCBL {}\ \BBA {} Andreotti%
}{%
Elbelrhiti%
\ \protect \BOthers {.}}{%
{\protect \APACyear {2005}}%
}]{%
Elbelrhiti}
\APACinsertmetastar {%
Elbelrhiti}%
\begin{APACrefauthors}%
Elbelrhiti, H.%
, Claudin, P.%
\BCBL {}\ \BBA {} Andreotti, B.%
\end{APACrefauthors}%
\unskip\
\newblock
\APACrefYearMonthDay{2005}{}{}.
\newblock
{\BBOQ}\APACrefatitle {Field evidence for surface-wave-induced instability of
  sand dunes} {Field evidence for surface-wave-induced instability of sand
  dunes}.{\BBCQ}
\newblock
\APACjournalVolNumPages{Nature}{437}{04058}{}.
\PrintBackRefs{\CurrentBib}

\bibitem [\protect \citeauthoryear {%
Endo%
, Taniguchi%
\BCBL {}\ \BBA {} Katsuki%
}{%
Endo%
\ \protect \BOthers {.}}{%
{\protect \APACyear {2004}}%
}]{%
Endo2}
\APACinsertmetastar {%
Endo2}%
\begin{APACrefauthors}%
Endo, N.%
, Taniguchi, K.%
\BCBL {}\ \BBA {} Katsuki, A.%
\end{APACrefauthors}%
\unskip\
\newblock
\APACrefYearMonthDay{2004}{}{}.
\newblock
{\BBOQ}\APACrefatitle {Observation of the whole process of interaction between
  barchans by flume experiments} {Observation of the whole process of
  interaction between barchans by flume experiments}.{\BBCQ}
\newblock
\APACjournalVolNumPages{Geophys. Res. Lett.}{31}{12}{}.
\PrintBackRefs{\CurrentBib}

\bibitem [\protect \citeauthoryear {%
Franklin%
}{%
Franklin%
}{%
{\protect \APACyear {2015}}%
}]{%
Franklin_12}
\APACinsertmetastar {%
Franklin_12}%
\begin{APACrefauthors}%
Franklin, E\BPBI M.%
\end{APACrefauthors}%
\unskip\
\newblock
\APACrefYearMonthDay{2015}{}{}.
\newblock
{\BBOQ}\APACrefatitle {Formation of sand ripples under a turbulent liquid flow}
  {Formation of sand ripples under a turbulent liquid flow}.{\BBCQ}
\newblock
\APACjournalVolNumPages{Appl. Math. Model.}{39}{23}{7390-7400}.
\newblock
\begin{APACrefURL}
  \url{https://www.sciencedirect.com/science/article/pii/S0307904X15001730}
  \end{APACrefURL}
\newblock
\begin{APACrefDOI} \doi{https://doi.org/10.1016/j.apm.2015.03.021}
  \end{APACrefDOI}
\PrintBackRefs{\CurrentBib}

\bibitem [\protect \citeauthoryear {%
Gay%
}{%
Gay%
}{%
{\protect \APACyear {1999}}%
}]{%
Gay}
\APACinsertmetastar {%
Gay}%
\begin{APACrefauthors}%
Gay, S\BPBI P.%
\end{APACrefauthors}%
\unskip\
\newblock
\APACrefYearMonthDay{1999}{}{}.
\newblock
{\BBOQ}\APACrefatitle {Observations regarding the movement of barchan sand
  dunes in the Nazca to Tanaca area of southern Peru} {Observations regarding
  the movement of barchan sand dunes in the nazca to tanaca area of southern
  peru}.{\BBCQ}
\newblock
\APACjournalVolNumPages{Geomorphology}{27}{3}{279 - 293}.
\PrintBackRefs{\CurrentBib}

\bibitem [\protect \citeauthoryear {%
G\'enois%
, du Pont%
, Hersen%
\BCBL {}\ \BBA {} Gr\'egoire%
}{%
G\'enois%
, du Pont%
\BCBL {}\ \protect \BOthers {.}}{%
{\protect \APACyear {2013}}%
}]{%
Genois2}
\APACinsertmetastar {%
Genois2}%
\begin{APACrefauthors}%
G\'enois, M.%
, du Pont, S\BPBI C.%
, Hersen, P.%
\BCBL {}\ \BBA {} Gr\'egoire, G.%
\end{APACrefauthors}%
\unskip\
\newblock
\APACrefYearMonthDay{2013}{}{}.
\newblock
{\BBOQ}\APACrefatitle {An agent-based model of dune interactions produces the
  emergence of patterns in deserts} {An agent-based model of dune interactions
  produces the emergence of patterns in deserts}.{\BBCQ}
\newblock
\APACjournalVolNumPages{Geophys. Res. Lett.}{40}{15}{3909-3914}.
\PrintBackRefs{\CurrentBib}

\bibitem [\protect \citeauthoryear {%
G\'enois%
, Hersen%
, du Pont%
\BCBL {}\ \BBA {} Gr\'egoire%
}{%
G\'enois%
, Hersen%
\BCBL {}\ \protect \BOthers {.}}{%
{\protect \APACyear {2013}}%
}]{%
Genois}
\APACinsertmetastar {%
Genois}%
\begin{APACrefauthors}%
G\'enois, M.%
, Hersen, P.%
, du Pont, S.%
\BCBL {}\ \BBA {} Gr\'egoire, G.%
\end{APACrefauthors}%
\unskip\
\newblock
\APACrefYearMonthDay{2013}{}{}.
\newblock
{\BBOQ}\APACrefatitle {Spatial structuring and size selection as collective
  behaviours in an agent-based model for barchan fields} {Spatial structuring
  and size selection as collective behaviours in an agent-based model for
  barchan fields}.{\BBCQ}
\newblock
\APACjournalVolNumPages{Eur. Phys. J. B}{86}{447}{}.
\PrintBackRefs{\CurrentBib}

\bibitem [\protect \citeauthoryear {%
Goniva%
, Kloss%
, Deen%
, Kuipers%
\BCBL {}\ \BBA {} Pirker%
}{%
Goniva%
\ \protect \BOthers {.}}{%
{\protect \APACyear {2012}}%
}]{%
Goniva}
\APACinsertmetastar {%
Goniva}%
\begin{APACrefauthors}%
Goniva, C.%
, Kloss, C.%
, Deen, N\BPBI G.%
, Kuipers, J\BPBI A\BPBI M.%
\BCBL {}\ \BBA {} Pirker, S.%
\end{APACrefauthors}%
\unskip\
\newblock
\APACrefYearMonthDay{2012}{}{}.
\newblock
{\BBOQ}\APACrefatitle {Influence of rolling friction on single spout fluidized
  bed simulation} {Influence of rolling friction on single spout fluidized bed
  simulation}.{\BBCQ}
\newblock
\APACjournalVolNumPages{Particuology}{10}{5}{582-591}.
\PrintBackRefs{\CurrentBib}

\bibitem [\protect \citeauthoryear {%
He%
, Lin%
, Zhang%
, Yang%
\BCBL {}\ \BBA {} Gao%
}{%
He%
\ \protect \BOthers {.}}{%
{\protect \APACyear {2023}}%
}]{%
He}
\APACinsertmetastar {%
He}%
\begin{APACrefauthors}%
He, N.%
, Lin, Y.%
, Zhang, Y.%
, Yang, B.%
\BCBL {}\ \BBA {} Gao, X.%
\end{APACrefauthors}%
\unskip\
\newblock
\APACrefYearMonthDay{2023}{10}{}.
\newblock
{\BBOQ}\APACrefatitle {{Self-stabilization of barchan dune chasing}}
  {{Self-stabilization of barchan dune chasing}}.{\BBCQ}
\newblock
\APACjournalVolNumPages{Phys. Fluids}{35}{10}{106609}.
\newblock
\begin{APACrefURL} \url{https://doi.org/10.1063/5.0169485} \end{APACrefURL}
\newblock
\begin{APACrefDOI} \doi{10.1063/5.0169485} \end{APACrefDOI}
\PrintBackRefs{\CurrentBib}

\bibitem [\protect \citeauthoryear {%
Herrmann%
\ \BBA {} Sauermann%
}{%
Herrmann%
\ \BBA {} Sauermann%
}{%
{\protect \APACyear {2000}}%
}]{%
Herrmann_Sauermann}
\APACinsertmetastar {%
Herrmann_Sauermann}%
\begin{APACrefauthors}%
Herrmann, H\BPBI J.%
\BCBT {}\ \BBA {} Sauermann, G.%
\end{APACrefauthors}%
\unskip\
\newblock
\APACrefYearMonthDay{2000}{}{}.
\newblock
{\BBOQ}\APACrefatitle {The shape of dunes} {The shape of dunes}.{\BBCQ}
\newblock
\APACjournalVolNumPages{Physica A (Amsterdam)}{283}{}{24-30}.
\PrintBackRefs{\CurrentBib}

\bibitem [\protect \citeauthoryear {%
Hersen%
}{%
Hersen%
}{%
{\protect \APACyear {2004}}%
}]{%
Hersen_3}
\APACinsertmetastar {%
Hersen_3}%
\begin{APACrefauthors}%
Hersen, P.%
\end{APACrefauthors}%
\unskip\
\newblock
\APACrefYearMonthDay{2004}{}{}.
\newblock
{\BBOQ}\APACrefatitle {On the crescentic shape of barchan dunes} {On the
  crescentic shape of barchan dunes}.{\BBCQ}
\newblock
\APACjournalVolNumPages{Eur. Phys. J. B}{37}{4}{507--514}.
\PrintBackRefs{\CurrentBib}

\bibitem [\protect \citeauthoryear {%
Hersen%
\ \protect \BOthers {.}}{%
Hersen%
\ \protect \BOthers {.}}{%
{\protect \APACyear {2004}}%
}]{%
Hersen_2}
\APACinsertmetastar {%
Hersen_2}%
\begin{APACrefauthors}%
Hersen, P.%
, Andersen, K\BPBI H.%
, Elbelrhiti, H.%
, Andreotti, B.%
, Claudin, P.%
\BCBL {}\ \BBA {} Douady, S.%
\end{APACrefauthors}%
\unskip\
\newblock
\APACrefYearMonthDay{2004}{Jan}{}.
\newblock
{\BBOQ}\APACrefatitle {Corridors of barchan dunes: Stability and size
  selection} {Corridors of barchan dunes: Stability and size selection}.{\BBCQ}
\newblock
\APACjournalVolNumPages{Phys. Rev. E}{69}{}{011304}.
\newblock
\begin{APACrefURL} \url{https://link.aps.org/doi/10.1103/PhysRevE.69.011304}
  \end{APACrefURL}
\newblock
\begin{APACrefDOI} \doi{10.1103/PhysRevE.69.011304} \end{APACrefDOI}
\PrintBackRefs{\CurrentBib}

\bibitem [\protect \citeauthoryear {%
Hersen%
\ \BBA {} Douady%
}{%
Hersen%
\ \BBA {} Douady%
}{%
{\protect \APACyear {2005}}%
}]{%
Hersen_5}
\APACinsertmetastar {%
Hersen_5}%
\begin{APACrefauthors}%
Hersen, P.%
\BCBT {}\ \BBA {} Douady, S.%
\end{APACrefauthors}%
\unskip\
\newblock
\APACrefYearMonthDay{2005}{}{}.
\newblock
{\BBOQ}\APACrefatitle {Collision of barchan dunes as a mechanism of size
  regulation} {Collision of barchan dunes as a mechanism of size
  regulation}.{\BBCQ}
\newblock
\APACjournalVolNumPages{Geophys. Res. Lett.}{32}{21}{}.
\PrintBackRefs{\CurrentBib}

\bibitem [\protect \citeauthoryear {%
Hugenholtz%
\ \BBA {} Barchyn%
}{%
Hugenholtz%
\ \BBA {} Barchyn%
}{%
{\protect \APACyear {2012}}%
}]{%
Hugenholtz}
\APACinsertmetastar {%
Hugenholtz}%
\begin{APACrefauthors}%
Hugenholtz, C\BPBI H.%
\BCBT {}\ \BBA {} Barchyn, T\BPBI E.%
\end{APACrefauthors}%
\unskip\
\newblock
\APACrefYearMonthDay{2012}{}{}.
\newblock
{\BBOQ}\APACrefatitle {Real barchan dune collisions and ejections} {Real
  barchan dune collisions and ejections}.{\BBCQ}
\newblock
\APACjournalVolNumPages{Geophys. Res. Lett.}{39}{2}{}.
\PrintBackRefs{\CurrentBib}

\bibitem [\protect \citeauthoryear {%
Katsuki%
, Kikuchi%
, Nishimori%
, Endo%
\BCBL {}\ \BBA {} Taniguchi%
}{%
Katsuki%
\ \protect \BOthers {.}}{%
{\protect \APACyear {2011}}%
}]{%
Katsuki}
\APACinsertmetastar {%
Katsuki}%
\begin{APACrefauthors}%
Katsuki, A.%
, Kikuchi, M.%
, Nishimori, H.%
, Endo, N.%
\BCBL {}\ \BBA {} Taniguchi, K.%
\end{APACrefauthors}%
\unskip\
\newblock
\APACrefYearMonthDay{2011}{}{}.
\newblock
{\BBOQ}\APACrefatitle {Cellular model for sand dunes with saltation, avalanche
  and strong erosion: collisional simulation of barchans} {Cellular model for
  sand dunes with saltation, avalanche and strong erosion: collisional
  simulation of barchans}.{\BBCQ}
\newblock
\APACjournalVolNumPages{Earth Surf. Process. Landforms}{36}{3}{372-382}.
\PrintBackRefs{\CurrentBib}

\bibitem [\protect \citeauthoryear {%
Kloss%
\ \BBA {} Goniva%
}{%
Kloss%
\ \BBA {} Goniva%
}{%
{\protect \APACyear {2010}}%
}]{%
Kloss}
\APACinsertmetastar {%
Kloss}%
\begin{APACrefauthors}%
Kloss, C.%
\BCBT {}\ \BBA {} Goniva, C.%
\end{APACrefauthors}%
\unskip\
\newblock
\APACrefYearMonthDay{2010}{}{}.
\newblock
{\BBOQ}\APACrefatitle {{LIGGGHTS}: a new open source discrete element
  simulation software} {{LIGGGHTS}: a new open source discrete element
  simulation software}.{\BBCQ}
\newblock
\BIn{} \APACrefbtitle {Proc. 5th Int. Conf. on Discrete Element Methods.}
  {Proc. 5th int. conf. on discrete element methods.}
\newblock
\APACaddressPublisher{London, UK}{}.
\PrintBackRefs{\CurrentBib}

\bibitem [\protect \citeauthoryear {%
Kocurek%
, Ewing%
\BCBL {}\ \BBA {} Mohrig%
}{%
Kocurek%
\ \protect \BOthers {.}}{%
{\protect \APACyear {2010}}%
}]{%
Kocurek}
\APACinsertmetastar {%
Kocurek}%
\begin{APACrefauthors}%
Kocurek, G.%
, Ewing, R\BPBI C.%
\BCBL {}\ \BBA {} Mohrig, D.%
\end{APACrefauthors}%
\unskip\
\newblock
\APACrefYearMonthDay{2010}{}{}.
\newblock
{\BBOQ}\APACrefatitle {How do bedform patterns arise? New views on the role of
  bedform interactions within a set of boundary conditions} {How do bedform
  patterns arise? new views on the role of bedform interactions within a set of
  boundary conditions}.{\BBCQ}
\newblock
\APACjournalVolNumPages{Earth Surf. Process. Landforms}{35}{1}{51-63}.
\PrintBackRefs{\CurrentBib}

\bibitem [\protect \citeauthoryear {%
Kroy%
, Sauermann%
\BCBL {}\ \BBA {} Herrmann%
}{%
Kroy%
\ \protect \BOthers {.}}{%
{\protect \APACyear {2002}}%
}]{%
Kroy_C}
\APACinsertmetastar {%
Kroy_C}%
\begin{APACrefauthors}%
Kroy, K.%
, Sauermann, G.%
\BCBL {}\ \BBA {} Herrmann, H\BPBI J.%
\end{APACrefauthors}%
\unskip\
\newblock
\APACrefYearMonthDay{2002}{Jan}{}.
\newblock
{\BBOQ}\APACrefatitle {Minimal Model for Sand Dunes} {Minimal model for sand
  dunes}.{\BBCQ}
\newblock
\APACjournalVolNumPages{Phys. Rev. Lett.}{88}{}{054301}.
\newblock
\begin{APACrefURL} \url{https://link.aps.org/doi/10.1103/PhysRevLett.88.054301}
  \end{APACrefURL}
\newblock
\begin{APACrefDOI} \doi{10.1103/PhysRevLett.88.054301} \end{APACrefDOI}
\PrintBackRefs{\CurrentBib}

\bibitem [\protect \citeauthoryear {%
A.~Lima%
, Sauermann%
, Herrmann%
\BCBL {}\ \BBA {} Kroy%
}{%
A.~Lima%
\ \protect \BOthers {.}}{%
{\protect \APACyear {2002}}%
}]{%
Lima}
\APACinsertmetastar {%
Lima}%
\begin{APACrefauthors}%
Lima, A.%
, Sauermann, G.%
, Herrmann, H.%
\BCBL {}\ \BBA {} Kroy, K.%
\end{APACrefauthors}%
\unskip\
\newblock
\APACrefYearMonthDay{2002}{}{}.
\newblock
{\BBOQ}\APACrefatitle {Modelling a dune field} {Modelling a dune field}.{\BBCQ}
\newblock
\APACjournalVolNumPages{Physica A}{310}{3}{487-500}.
\newblock
\begin{APACrefURL}
  \url{https://www.sciencedirect.com/science/article/pii/S0378437102005460}
  \end{APACrefURL}
\PrintBackRefs{\CurrentBib}

\bibitem [\protect \citeauthoryear {%
N\BPBI C.~Lima%
, Assis%
, Alvarez%
\BCBL {}\ \BBA {} Franklin%
}{%
N\BPBI C.~Lima%
\ \protect \BOthers {.}}{%
{\protect \APACyear {2022}}%
}]{%
Lima2}
\APACinsertmetastar {%
Lima2}%
\begin{APACrefauthors}%
Lima, N\BPBI C.%
, Assis, W\BPBI R.%
, Alvarez, C\BPBI A.%
\BCBL {}\ \BBA {} Franklin, E\BPBI M.%
\end{APACrefauthors}%
\unskip\
\newblock
\APACrefYearMonthDay{2022}{}{}.
\newblock
{\BBOQ}\APACrefatitle {Grain-scale computations of barchan dunes} {Grain-scale
  computations of barchan dunes}.{\BBCQ}
\newblock
\APACjournalVolNumPages{Phys. Fluids}{34}{12}{123320}.
\newblock
\begin{APACrefURL} \url{https://doi.org/10.1063/5.0121810} \end{APACrefURL}
\newblock
\begin{APACrefDOI} \doi{10.1063/5.0121810} \end{APACrefDOI}
\PrintBackRefs{\CurrentBib}

\bibitem [\protect \citeauthoryear {%
N\BPBI C.~Lima%
, Assis%
\BCBL {}\ \BBA {} Franklin%
}{%
N\BPBI C.~Lima%
\ \protect \BOthers {.}}{%
{\protect \APACyear {2024}}%
}]{%
Supplemental2}
\APACinsertmetastar {%
Supplemental2}%
\begin{APACrefauthors}%
Lima, N\BPBI C.%
, Assis, W\BPBI R.%
\BCBL {}\ \BBA {} Franklin, E\BPBI M.%
\end{APACrefauthors}%
\unskip\
\newblock
\APACrefYearMonthDay{2024}{}{}.
\newblock
{\BBOQ}\APACrefatitle {Numerical Simulation of Barchan-barchan repulsion
  [{D}ataset][{S}oftware]} {Numerical simulation of barchan-barchan repulsion
  [{D}ataset][{S}oftware]}.{\BBCQ}
\newblock
\APACjournalVolNumPages{Mendeley Data,
  http://dx.doi.org/10.17632/ypkgwjfr4r.1}{}{}{}.
\newblock
\begin{APACrefDOI} \doi{10.17632/ypkgwjfr4r.1} \end{APACrefDOI}
\PrintBackRefs{\CurrentBib}

\bibitem [\protect \citeauthoryear {%
Liu%
, Liu%
, Fu%
\BCBL {}\ \BBA {} G.%
}{%
Liu%
\ \protect \BOthers {.}}{%
{\protect \APACyear {2016}}%
}]{%
Liu}
\APACinsertmetastar {%
Liu}%
\begin{APACrefauthors}%
Liu, D.%
, Liu, X.%
, Fu, X.%
\BCBL {}\ \BBA {} G., W.%
\end{APACrefauthors}%
\unskip\
\newblock
\APACrefYearMonthDay{2016}{}{}.
\newblock
{\BBOQ}\APACrefatitle {Quantification of the bed load effects on turbulent
  open-channel flows} {Quantification of the bed load effects on turbulent
  open-channel flows}.{\BBCQ}
\newblock
\APACjournalVolNumPages{J. Geophys. Res. Earth Surf.}{121}{}{767-789}.
\PrintBackRefs{\CurrentBib}

\bibitem [\protect \citeauthoryear {%
Narteau%
, Zhang%
, Rozier%
\BCBL {}\ \BBA {} Claudin%
}{%
Narteau%
\ \protect \BOthers {.}}{%
{\protect \APACyear {2009}}%
}]{%
Narteau}
\APACinsertmetastar {%
Narteau}%
\begin{APACrefauthors}%
Narteau, C.%
, Zhang, D.%
, Rozier, O.%
\BCBL {}\ \BBA {} Claudin, P.%
\end{APACrefauthors}%
\unskip\
\newblock
\APACrefYearMonthDay{2009}{}{}.
\newblock
{\BBOQ}\APACrefatitle {Setting the length and time scales of a cellular
  automaton dune model from the analysis of superimposed bed forms} {Setting
  the length and time scales of a cellular automaton dune model from the
  analysis of superimposed bed forms}.{\BBCQ}
\newblock
\APACjournalVolNumPages{J. Geophys. Res.: Earth Surf.}{114}{F3}{}.
\PrintBackRefs{\CurrentBib}

\bibitem [\protect \citeauthoryear {%
Norris%
\ \BBA {} Norris%
}{%
Norris%
\ \BBA {} Norris%
}{%
{\protect \APACyear {1961}}%
}]{%
Norris}
\APACinsertmetastar {%
Norris}%
\begin{APACrefauthors}%
Norris, R\BPBI M.%
\BCBT {}\ \BBA {} Norris, K\BPBI S.%
\end{APACrefauthors}%
\unskip\
\newblock
\APACrefYearMonthDay{1961}{}{}.
\newblock
{\BBOQ}\APACrefatitle {Algodones {D}unes of {S}outheastern {C}alifornia}
  {Algodones {D}unes of {S}outheastern {C}alifornia}.{\BBCQ}
\newblock
\APACjournalVolNumPages{GSA Bulletin}{72}{4}{605-619}.
\PrintBackRefs{\CurrentBib}

\bibitem [\protect \citeauthoryear {%
E.~Parteli%
\ \BBA {} Herrmann%
}{%
E.~Parteli%
\ \BBA {} Herrmann%
}{%
{\protect \APACyear {2003}}%
}]{%
Partelli6}
\APACinsertmetastar {%
Partelli6}%
\begin{APACrefauthors}%
Parteli, E.%
\BCBT {}\ \BBA {} Herrmann, H.%
\end{APACrefauthors}%
\unskip\
\newblock
\APACrefYearMonthDay{2003}{}{}.
\newblock
{\BBOQ}\APACrefatitle {A simple model for a transverse dune field} {A simple
  model for a transverse dune field}.{\BBCQ}
\newblock
\APACjournalVolNumPages{Physica A}{327}{3}{554-562}.
\PrintBackRefs{\CurrentBib}

\bibitem [\protect \citeauthoryear {%
E\BPBI J\BPBI R.~Parteli%
\ \protect \BOthers {.}}{%
E\BPBI J\BPBI R.~Parteli%
\ \protect \BOthers {.}}{%
{\protect \APACyear {2014}}%
}]{%
Parteli4}
\APACinsertmetastar {%
Parteli4}%
\begin{APACrefauthors}%
Parteli, E\BPBI J\BPBI R.%
, Dur{\'a}n, O.%
, Bourke, M\BPBI C.%
, Tsoar, H.%
, P{\"o}schel, T.%
\BCBL {}\ \BBA {} Herrmann, H.%
\end{APACrefauthors}%
\unskip\
\newblock
\APACrefYearMonthDay{2014}{}{}.
\newblock
{\BBOQ}\APACrefatitle {Origins of barchan dune asymmetry: Insights from
  numerical simulations} {Origins of barchan dune asymmetry: Insights from
  numerical simulations}.{\BBCQ}
\newblock
\APACjournalVolNumPages{Aeol. Res.}{12}{}{121--133}.
\PrintBackRefs{\CurrentBib}

\bibitem [\protect \citeauthoryear {%
E\BPBI J\BPBI R.~Parteli%
\ \BBA {} Herrmann%
}{%
E\BPBI J\BPBI R.~Parteli%
\ \BBA {} Herrmann%
}{%
{\protect \APACyear {2007}}%
}]{%
Parteli2}
\APACinsertmetastar {%
Parteli2}%
\begin{APACrefauthors}%
Parteli, E\BPBI J\BPBI R.%
\BCBT {}\ \BBA {} Herrmann, H\BPBI J.%
\end{APACrefauthors}%
\unskip\
\newblock
\APACrefYearMonthDay{2007}{Oct}{}.
\newblock
{\BBOQ}\APACrefatitle {Dune formation on the present Mars} {Dune formation on
  the present mars}.{\BBCQ}
\newblock
\APACjournalVolNumPages{Phys. Rev. E}{76}{}{041307}.
\newblock
\begin{APACrefURL} \url{https://link.aps.org/doi/10.1103/PhysRevE.76.041307}
  \end{APACrefURL}
\newblock
\begin{APACrefDOI} \doi{10.1103/PhysRevE.76.041307} \end{APACrefDOI}
\PrintBackRefs{\CurrentBib}

\bibitem [\protect \citeauthoryear {%
Robson%
\ \BBA {} Baas%
}{%
Robson%
\ \BBA {} Baas%
}{%
{\protect \APACyear {2023}}%
}]{%
Robson}
\APACinsertmetastar {%
Robson}%
\begin{APACrefauthors}%
Robson, D\BPBI T.%
\BCBT {}\ \BBA {} Baas, A\BPBI C\BPBI W.%
\end{APACrefauthors}%
\unskip\
\newblock
\APACrefYearMonthDay{2023}{}{}.
\newblock
{\BBOQ}\APACrefatitle {Barchan swarm dynamics from a Two-Flank Agent-Based
  Model} {Barchan swarm dynamics from a two-flank agent-based model}.{\BBCQ}
\newblock
\APACjournalVolNumPages{EGUsphere}{2023}{}{1--39}.
\newblock
\begin{APACrefURL}
  \url{https://egusphere.copernicus.org/preprints/2023/egusphere-2023-2900/}
  \end{APACrefURL}
\newblock
\begin{APACrefDOI} \doi{10.5194/egusphere-2023-2900} \end{APACrefDOI}
\PrintBackRefs{\CurrentBib}

\bibitem [\protect \citeauthoryear {%
Sauermann%
, Kroy%
\BCBL {}\ \BBA {} Herrmann%
}{%
Sauermann%
\ \protect \BOthers {.}}{%
{\protect \APACyear {2001}}%
}]{%
Sauermann_4}
\APACinsertmetastar {%
Sauermann_4}%
\begin{APACrefauthors}%
Sauermann, G.%
, Kroy, K.%
\BCBL {}\ \BBA {} Herrmann, H\BPBI J.%
\end{APACrefauthors}%
\unskip\
\newblock
\APACrefYearMonthDay{2001}{Aug}{}.
\newblock
{\BBOQ}\APACrefatitle {Continuum saltation model for sand dunes} {Continuum
  saltation model for sand dunes}.{\BBCQ}
\newblock
\APACjournalVolNumPages{Phys. Rev. E}{64}{}{031305}.
\newblock
\begin{APACrefURL} \url{https://link.aps.org/doi/10.1103/PhysRevE.64.031305}
  \end{APACrefURL}
\newblock
\begin{APACrefDOI} \doi{10.1103/PhysRevE.64.031305} \end{APACrefDOI}
\PrintBackRefs{\CurrentBib}

\bibitem [\protect \citeauthoryear {%
Schw{\"a}mmle%
\ \BBA {} Herrmann%
}{%
Schw{\"a}mmle%
\ \BBA {} Herrmann%
}{%
{\protect \APACyear {2003}}%
}]{%
Schwammle2}
\APACinsertmetastar {%
Schwammle2}%
\begin{APACrefauthors}%
Schw{\"a}mmle, V.%
\BCBT {}\ \BBA {} Herrmann, H\BPBI J.%
\end{APACrefauthors}%
\unskip\
\newblock
\APACrefYearMonthDay{2003}{}{}.
\newblock
{\BBOQ}\APACrefatitle {Solitary wave behaviour of sand dunes} {Solitary wave
  behaviour of sand dunes}.{\BBCQ}
\newblock
\APACjournalVolNumPages{Nature}{426}{}{619-620}.
\PrintBackRefs{\CurrentBib}

\bibitem [\protect \citeauthoryear {%
Schw{\"a}mmle%
\ \BBA {} Herrmann%
}{%
Schw{\"a}mmle%
\ \BBA {} Herrmann%
}{%
{\protect \APACyear {2005}}%
}]{%
Schwammle}
\APACinsertmetastar {%
Schwammle}%
\begin{APACrefauthors}%
Schw{\"a}mmle, V.%
\BCBT {}\ \BBA {} Herrmann, H\BPBI J.%
\end{APACrefauthors}%
\unskip\
\newblock
\APACrefYearMonthDay{2005}{}{}.
\newblock
{\BBOQ}\APACrefatitle {A model of Barchan dunes including lateral shear stress}
  {A model of barchan dunes including lateral shear stress}.{\BBCQ}
\newblock
\APACjournalVolNumPages{Eur. Phys. J. E}{16}{1}{57-65}.
\PrintBackRefs{\CurrentBib}

\bibitem [\protect \citeauthoryear {%
Tsuji%
, Kawaguchi%
\BCBL {}\ \BBA {} Tanaka%
}{%
Tsuji%
\ \protect \BOthers {.}}{%
{\protect \APACyear {1993}}%
}]{%
Tsuji2}
\APACinsertmetastar {%
Tsuji2}%
\begin{APACrefauthors}%
Tsuji, Y.%
, Kawaguchi, T.%
\BCBL {}\ \BBA {} Tanaka, T.%
\end{APACrefauthors}%
\unskip\
\newblock
\APACrefYearMonthDay{1993}{}{}.
\newblock
{\BBOQ}\APACrefatitle {Discrete particle simulation of two-dimensional
  fluidized bed} {Discrete particle simulation of two-dimensional fluidized
  bed}.{\BBCQ}
\newblock
\APACjournalVolNumPages{Powder Technology}{77}{1}{79-87}.
\PrintBackRefs{\CurrentBib}

\bibitem [\protect \citeauthoryear {%
Tsuji%
, Tanaka%
\BCBL {}\ \BBA {} Ishida%
}{%
Tsuji%
\ \protect \BOthers {.}}{%
{\protect \APACyear {1992}}%
}]{%
Tsuji}
\APACinsertmetastar {%
Tsuji}%
\begin{APACrefauthors}%
Tsuji, Y.%
, Tanaka, T.%
\BCBL {}\ \BBA {} Ishida, T.%
\end{APACrefauthors}%
\unskip\
\newblock
\APACrefYearMonthDay{1992}{}{}.
\newblock
{\BBOQ}\APACrefatitle {Lagrangian numerical simulation of plug flow of
  cohesionless particles in a horizontal pipe} {Lagrangian numerical simulation
  of plug flow of cohesionless particles in a horizontal pipe}.{\BBCQ}
\newblock
\APACjournalVolNumPages{Powder Technology}{71}{3}{239-250}.
\PrintBackRefs{\CurrentBib}

\bibitem [\protect \citeauthoryear {%
Vermeesch%
}{%
Vermeesch%
}{%
{\protect \APACyear {2011}}%
}]{%
Vermeesch}
\APACinsertmetastar {%
Vermeesch}%
\begin{APACrefauthors}%
Vermeesch, P.%
\end{APACrefauthors}%
\unskip\
\newblock
\APACrefYearMonthDay{2011}{}{}.
\newblock
{\BBOQ}\APACrefatitle {Solitary wave behavior in sand dunes observed from
  space} {Solitary wave behavior in sand dunes observed from space}.{\BBCQ}
\newblock
\APACjournalVolNumPages{Geophys. Res. Lett.}{38}{22}{}.
\PrintBackRefs{\CurrentBib}

\bibitem [\protect \citeauthoryear {%
Wenzel%
\ \BBA {} Franklin%
}{%
Wenzel%
\ \BBA {} Franklin%
}{%
{\protect \APACyear {2019}}%
}]{%
Wenzel}
\APACinsertmetastar {%
Wenzel}%
\begin{APACrefauthors}%
Wenzel, J\BPBI L.%
\BCBT {}\ \BBA {} Franklin, E\BPBI M.%
\end{APACrefauthors}%
\unskip\
\newblock
\APACrefYearMonthDay{2019}{}{}.
\newblock
{\BBOQ}\APACrefatitle {Velocity fields and particle trajectories for bed load
  over subaqueous barchan dunes} {Velocity fields and particle trajectories for
  bed load over subaqueous barchan dunes}.{\BBCQ}
\newblock
\APACjournalVolNumPages{Granular Matter}{21}{}{321-334}.
\PrintBackRefs{\CurrentBib}

\bibitem [\protect \citeauthoryear {%
Zhang%
, Yang%
, Rozier%
\BCBL {}\ \BBA {} Narteau%
}{%
Zhang%
\ \protect \BOthers {.}}{%
{\protect \APACyear {2014}}%
}]{%
Zhang_D}
\APACinsertmetastar {%
Zhang_D}%
\begin{APACrefauthors}%
Zhang, D.%
, Yang, X.%
, Rozier, O.%
\BCBL {}\ \BBA {} Narteau, C.%
\end{APACrefauthors}%
\unskip\
\newblock
\APACrefYearMonthDay{2014}{}{}.
\newblock
{\BBOQ}\APACrefatitle {Mean sediment residence time in barchan dunes} {Mean
  sediment residence time in barchan dunes}.{\BBCQ}
\newblock
\APACjournalVolNumPages{J. Geophys. Res.: Earth Surf.}{119}{3}{451--463}.
\PrintBackRefs{\CurrentBib}

\bibitem [\protect \citeauthoryear {%
X.~Zhou%
, Wang%
\BCBL {}\ \BBA {} Yang%
}{%
X.~Zhou%
\ \protect \BOthers {.}}{%
{\protect \APACyear {2019}}%
}]{%
Zhou2}
\APACinsertmetastar {%
Zhou2}%
\begin{APACrefauthors}%
Zhou, X.%
, Wang, Y.%
\BCBL {}\ \BBA {} Yang, B.%
\end{APACrefauthors}%
\unskip\
\newblock
\APACrefYearMonthDay{2019}{}{}.
\newblock
{\BBOQ}\APACrefatitle {Three-dimensional numerical simulations of barchan dune
  interactions in unidirectional flow} {Three-dimensional numerical simulations
  of barchan dune interactions in unidirectional flow}.{\BBCQ}
\newblock
\APACjournalVolNumPages{Particul. Sci. Technol.}{37}{7}{835-842}.
\PrintBackRefs{\CurrentBib}

\bibitem [\protect \citeauthoryear {%
Z\BPBI Y.~Zhou%
, Kuang%
, Chu%
\BCBL {}\ \BBA {} Yu%
}{%
Z\BPBI Y.~Zhou%
\ \protect \BOthers {.}}{%
{\protect \APACyear {2010}}%
}]{%
Zhou}
\APACinsertmetastar {%
Zhou}%
\begin{APACrefauthors}%
Zhou, Z\BPBI Y.%
, Kuang, S\BPBI B.%
, Chu, K\BPBI W.%
\BCBL {}\ \BBA {} Yu, A\BPBI B.%
\end{APACrefauthors}%
\unskip\
\newblock
\APACrefYearMonthDay{2010}{}{}.
\newblock
{\BBOQ}\APACrefatitle {Discrete particle simulation of particle–fluid flow:
  model formulations and their applicability} {Discrete particle simulation of
  particle–fluid flow: model formulations and their applicability}.{\BBCQ}
\newblock
\APACjournalVolNumPages{J. Fluid Mech.}{661}{}{482–510}.
\PrintBackRefs{\CurrentBib}

\end{thebibliography}

\end{document}